\documentclass[prb,aps,twocolumn,showpacs]{revtex4}
\usepackage{graphicx}
\usepackage{bm}
\def\ket#1{|#1\rangle }
\def\bra#1{\langle#1 | }

\def\punkt{\;\; .}
\def\komma{\;\; ,}
\def\expect#1{\langle#1 \rangle}
\def\mat#1{\underline{\underline{{\bf #1 }}}}
\def\w{\omega}
\def\H{{\cal H}}
\def\e{\epsilon}

\begin{document}

\title{Spin Precession and Real Time Dynamics in the Kondo Model:
       A Time-Dependent Numerical Renormalization-Group Study}

\smallskip

\author{Frithjof B.~Anders$^1$ and Avraham Schiller$^2$}
\affiliation{$^1$Department of Physics, Universit\"at Bremen,
                 P.O. Box 330 440, D-28334 Bremen, Germany\\
             $^2$Racah Institute of Physics, The Hebrew University,
                 Jerusalem 91904, Israel}
\date{\today}

\begin{abstract}
A detailed derivation of the recently proposed
time-dependent numerical renormalization-group (TD-NRG)
approach to nonequilibrium dynamics in quantum impurity
systems is presented. We demonstrate that the method
is suitable for fermionic as well as bosonic baths. A
comparison with exact analytical results for the charge
relaxation in the resonant-level model and for dephasing
in the spin-boson model establishes the accuracy of the
method. The real-time dynamics of a single spin coupled
to both types of baths is investigated. We use the TD-NRG
to calculate the spin relaxation and spin precession of
a single Kondo impurity. The short- and long-time dynamics
is studied as a function of temperature in the ferromagnetic
and antiferromagnetic regimes. The short-time dynamics
agrees very well with analytical results obtained at
second order in the exchange coupling $J$. In the
ferromagnetic regime, the long-time spin decay is described
by the scaling variable $x = 2\rho_F J(T) T t$. In the
antiferromagnetic regime it is governed for $T < T_K$
by the Kondo time scale $1/T_K$. Here $\rho_F$ is the
conduction-electron density of states and $T_K$ is the
Kondo temperature. Results for spin precession are
obtained by rotating the external magnetic field from
the $x$ axis to the $z$ axis.
\end{abstract}


\maketitle


\section{Introduction}

The decoherence and relaxation of an impurity spin is a classic
problem in condensed mater physics. More then 30 years ago,
Langreth and Wilkins developed a theory of spin resonance in dilute
magnetic alloys.~\cite{LangrethWilkins1972} The principal objective
was to derive Bloch-like equations for the paramagnetic resonance
using the Kadanoff-Baym~\cite{KadanoffBaym62} and Keldysh~\cite{Keldysh65}
techniques to nonequilibrium. This included the description
of spin precession and spin relaxation of a finite concentration
of magnetic impurities weakly coupled to a  metallic host.
The recent advent of quantum-dot devices and their potential
application to quantum computing has renewed interest in the spin
dynamics of a {\em single} impurity. In contrast to real magnetic
impurities, quantum dots can be controlled in exquisite detail,
and can be tuned at will from weak coupling to the Kondo regime.
Perturbative approaches, such as that of Langreth and
Wilkins,~\cite{LangrethWilkins1972} fail to describe the strong
correlation that develop in latter regime. This calls for new
theoretical approaches to nonequilibrium dynamics, suitable for
treating  the strongly correlated state.

In this work, we focus on the dynamics of a single spin coupled
to an infinitely large environment of noninteracting particles, in
order to investigate spin-relaxation phenomena at all temperatures
and coupling regimes. Relaxation of a single spin is the simplest
example for real-time dynamics in a quantum impurity system. Other
examples might be qubits, coupled quantum dots, or biological
donor-acceptor molecules. In such systems one is generically
interested in the dynamics of a finite subsystem interacting with
an infinitely large environment. The nature of the environment
may vary from one realization to another. It can consist of a
fermionic bath, as in single-electron and single-molecule
transistors, or a bosonic bath, as in the spin-boson model.
In certain cases a combined fermionic-bosonic bath might be
in order.

The immense difficulty of treating the real-time dynamics
of quantum impurity systems stems from the need to track
the full time evolution of the density operator of the entire
system --- environment plus impurity. The Kadanoff-Baym
and Keldysh techniques~\cite{Keldysh65,KadanoffBaym62}  
provide an elegant platform for perturbative expansions of the density
operator. In general, however, perturbative approaches are plagued by
the infra-red divergences caused by degeneracies on the impurity,
making them inadequate for tackling quantum impurities. Here
we take an alternative approach to the real-time dynamics based on
Wilson's numerical renormalization-group (NRG) method.~\cite{Wilson75}

The NRG is a very powerful tool for accurately calculating
equilibrium properties of arbitrarily complex quantum impurities.
Originally developed for treating the single-channel Kondo
Hamiltonian,~\cite{Wilson75} this nonperturbative approach
was successfully extended to the Anderson impurity model
(symmetric~\cite{KrishWilWilson80a} and
asymmetric~\cite{KrishWilWilson80b}), the two-channel Kondo
Hamiltonian,~\cite{Cragg_et_al,PangCox91} different two-impurity
clusters,~\cite{Jones_et_al_1987,Jones_et_al_1988,Sakai_et_al_1990,
Sakai_et_al_1992a,Sakai_et_al_1992b,IJW92} and a host of related
zero-dimensional problems. Recently, we extended the approach to
a certain class of time-dependent problems where a sudden
perturbation is applied to the impurity at time
$t = 0$.~\cite{AndersSchiller2005} Similar to the equilibrium NRG,
the time-dependent NRG (TD-NRG) can be applied to all coupling
strengths and is not confined to the weak-coupling regime. It is
capable of accessing all time scales, including arbitrary long as
well as arbitrary short times. These appealing properties make
the TD-NRG a powerful new approach for studying nonequilibrium
dynamics in quantum-impurity systems. 

In this work, we present the complete details of the TD-NRG,
and apply it to the spin dynamics of the Kondo model. A
comprehensive analysis is presented for the spin relaxation
and spin precession as a function of temperature, magnetic field
and coupling strength. Both ferromagnetic and antiferromagnetic
couplings are considered. In the Kondo regime, spin dynamics is
governed by the Kondo time scale $t_K = 1/T_K$, where $T_K$ is
the Kondo temperature of the system. In the ferromagnetic
regime, the underlying time scale is strongly dependent on
temperature, reflecting the fact that the effective equilibrium
coupling flows to zero with decreasing temperature.

\subsection{Preliminaries}

The idea of applying the NRG to nonequilibrium dynamics dates
back to the work of Costi.~\cite{Costi97} In the spirit of the
equilibrium NRG, Costi evaluated the nonequilibrium spectral
functions using an individual Wilson shell for each frequency
interval. However, as already recognized by Costi,~\cite{Costi97}
a full multiple-shell evaluation is ultimately required for
the correct description of nonequilibrium dynamics. The physical
reason is simple: When expanded in terms of the eigenstates of
the perturbed Hamiltonian, the initial state of the unperturbed
system contains contributions from all energy scales. This calls
for an adequate coupling of low- and high-energy scales absent
in the equilibrium NRG.

To circumvent this problem, we construct a complete basis set
of the Wilson chain using the NRG eigenstates. Implementing a
suitable resummation procedure, we track all states that
contribute to the time evolution of the observables of
interest.~\cite{AndersSchiller2005} Explicitly, we have shown
that the time evolution of a general local operator $\hat O$
can be written as
\begin{eqnarray}
\langle \hat{O} \rangle (t) &=&
        \sum_{m}^{N}\sum_{r,s}^{trun} \;
        e^{i(E_{r}^m -E_{s}^m)t}
        O_{r,s}^m \rho^{\rm red}_{s,r}(m) \; ,
\label{eqn:time-evolution-intro} 
\end{eqnarray}
where $E_{r}^m$ and $E_{s}^m$ are the dimension-full NRG
eigen-energies of the perturbed Hamiltonian at iteration
$m \le N$, $O_{r,s}^m$ is the matrix representation of
$\hat O$ at that iteration, and $\rho^{\rm red}_{s,r}(m)$
is the reduced density matrix~\cite{Feynman72} in the basis
of perturbed Hamiltonian. The restrictive sum over $r$
and $s$ requires that at least one of these states is
discarded at iteration $m$.
In context of the equilibrium NRG, the
reduced density matrix has been already used for the calculation of
spectral functions.~\cite{Hofstetter2000}

Equation~(\ref{eqn:time-evolution-intro}) is the centerpiece
of the TD-NRG approach and will be heavily using throughout
this paper. Below we present a detailed discussion of its
derivation and implementation. Here we only wish to emphasize
the following points.
(i) The reduced density matrix occurs naturally in our
    formulation.
(ii) It does not follow a unitary time evolution and, therefore,
     contains information on dissipation and  decoherence. 
(iii) Equation~(\ref{eqn:time-evolution-intro}) arises from
      summation over the {\em complete} many-body Fock space
      of the Wilson chain. No truncation of states is involved.

\subsection{Plan of the Paper}

As implied above, this paper has two main objectives: an indepth
presentation of the TD-NRG and a comprehensive analysis of the
spin relaxation for a single spin coupled to a conduction band.
Accordingly, the remainder of the paper divides into two major
parts.
 
The following three sections are devoted to an extensive
exposition of the TD-NRG. This includes a detailed derivation of
the approach, along with benchmark results for both a fermionic
and a bosonic bath. Section~\ref{sec:td-nrg-theory-general}
contains a general formulation of TD-NRG for any initial
density operator $\hat{\rho}_0$ of the system. In particular,
Eq.~(\ref{eqn:time-evolution-intro}) is derived and an explicit
definition of the reducted density matrix $\rho^{\rm red}(m)$ is
given. In Sec.~\ref{sec:reduced-density-matrix}, we discuss how
the reduced density matrix is calculated in practice in the
generic case where $\hat{\rho}_0$ is the equilibrium density
operator for the unperturbed system. The complete TD-NRG
algorithm is outlined in turn in Sec.~\ref{sec:td-nrg-algo}.
In sections~\ref{sec:rlm} and \ref{sec:spin-boson}, we compare
the TD-NRG to exact results for two distinct models: the
resonant-level model and a restricted version of the spin-boson
model. These two test cases allow us to establish the accuracy
of the TD-NRG on all time-scales, and to investigate how
well a discretized finite-size system can be used to mimic
nonequilibrium dynamics in a continuous bath.

Following this detailed exposition of the TD-NRG, we proceed
in sections~\ref{sec:kondo-model-analytics} and
\ref{sec:kondo-model-td-nrg} to the spin dynamics in the Kondo
model. In Sec.~\ref{sec:kondo-model-analytics}, we introduce
the model and present some basic analytical considerations.
These include an explicit calculation of the short-time
response up to second order in the dimensionless exchange
coupling. Numerical results for arbitrarily long time scale
are covered in turn in Sec.~\ref{sec:kondo-model-td-nrg},
both for the ferromagnetic and antiferromagnetic regimes.
Different configurations of the magnetic field are also
considered. Through comparison with the analytical results
of Sec.~\ref{sec:kondo-model-analytics} we are able to
establish the accuracy of the TD-NRG results at short time
scales. The true power of the numerical solution lies,
however, in the long-time behavior not accessible by
perturbative techniques.

\section{Theoretical formulation}
\label{sec:td-nrg-theory}

\subsection{Goal of the method}

In the present section we formulate the TD-NRG approach
in general terms, before turning to concrete examples
in sections~\ref{sec:rlm}, \ref{sec:spin-boson}, and
\ref{sec:kondo-model-td-nrg}. Hereafter we assume that the
initial density operator at time $t = 0$ is known, when a
sudden perturbation $\Delta {\cal H}$ is switch on. For
$t > 0$, the system evolves in time with the full Hamiltonian
${\cal H}^f = {\cal H}^i + \Delta {\cal H}$, where ${\cal H}^i$
is the Hamiltonian of the unperturbed system at time $t < 0$.
Denoting the initial density operator by $\hat{\rho}_0$,
the expectation value of any operator $\hat{O}$ at time
$t \ge 0$ is given by
\begin{eqnarray}
O(t) &=& {\rm Tr}\{ \hat{\rho}(t) \hat O \} \; ,
\label{eqn:time-evolution-general}
\end{eqnarray}
where $\hat{\rho}(t)$ is the corresponding time-dependent
density operator.

Our first goal is to show that expectation value $O(t)$ of
any local operator $\hat{O}$ can be rewritten in terms of a
sum over the contributions of a sequence of reduced density
matrices $\rho^{\rm red}(t)$.~\cite{Feynman72,AndersSchiller2005}
The reduced density matrices are generated by systematically
tracing out all environment degrees of freedom. Hence the
unitary time evolution of the density matrix of the system as
a whole is lost, giving rise to dissipation and decoherence.
Although our derivation makes explicit reference to the NRG
procedure, it might be applicable to other methods as well. 

Two key ingredients underlie our approach: (i) The identification
of a {\em complete basis set} of the many-body Fock space of
the Wilson chain based on the NRG eigenstates at the different
iterations; (ii) Expectation values are obtained by explicitly
tracing over this complete basis set. This marks a significant
departure from the conventional renormalization-group concept,
where the relevant physical information is contained in the
{\em kept states}. Here, the {\em discarded} states are
solemnly responsible for the time evolution. This also
solves the problem of overcounting excitations encountered
by Costi:~\cite{Costi97} Each excitation contributes only
once to expectation values, at that iteration where the
corresponding state is discarded. For the case where
$\hat{\rho}_0$ either projects onto a single state
or represents the full equilibrium density operator
of the unperturbed Hamiltonian, we are able give a closed
analytical expression for the reduced density matrices.

\subsection{Real-time evolution in quantum impurity systems} 
\label{sec:td-nrg-theory-general}

The Hamiltonian of a quantum impurity system is generally
given by
\begin{eqnarray}
{\cal H} = {\cal H}_{\rm bath} + {\cal H}_{\rm imp} +
           {\cal H}_{\rm mix} \; ,
\end{eqnarray}
where ${\cal H}_{\rm bath}$ models the continuous bath,
${\cal H}_{\rm imp}$ represents the decoupled impurity, and
${\cal H}_{\rm mix}$ describes the coupling between the two
subsystems. The entire system is characterized at time
$t = 0$ by an arbitrary density matrix $\hat{\rho}_0$, when
a sudden, time-independent perturbation $\Delta {\cal H}$
is switched on: ${\cal H}(t \ge 0) = {\cal H}^{i} +
\Delta {\cal H} \equiv {\cal H}^{f}$. For $t \ge 0 $, the
density operator evolves according to
\begin{equation}
\hat{\rho}(t) = e^{-i {\cal H}^f t} \hat{\rho}_0
                e^{i {\cal H}^f t} \; .
\label{rho-of-t}
\end{equation}

In equilibrium, one can solve such a quantum impurity system
very accurately using the NRG. At the heart of this approach is
a logarithmic discretization of the continuous bath, controlled
by the discretization parameter $\Lambda > 1$.~\cite{Wilson75}
The continuum limit is recovered for $\Lambda \to 1$. Using
an appropriate unitary transformation,~\cite{Wilson75} the
Hamiltonian is mapped onto a semi-infinite chain, with the
impurity coupled to the open end. The $N$th link along the
chain represents an exponentially decreasing energy scale:
$D_N \sim \Lambda^{-N/2}$ for a fermionic bath~\cite{Wilson75}
and $D_N \sim \Lambda^{-N}$ for a bosonic
bath.~\cite{BullaBoson2003} Using this hierarchy of scales,
the sequence of finite-size Hamiltonians ${\cal H}_N$ for the
$N$-site chain~\footnote{We use the standard notation, by
which the $N$-site chain contains the impurity degrees of
freedom, as well as the first $N + 1$ Wilson shells (labelled
by $n = 0, \cdots, N$). In contrary to the standard NRG notations we
consider $\H_N$ as dimensionfull Hamiltonian in order to emphasize the
importance of all energy shells for the real time evolution.}
is solved iteratively, discarding the high-energy states at
the conclusion of each step to maintain a manageable number
of states. The reduced basis set of ${\cal H}_N$ so obtained
is expected to faithfully describe the spectrum of the full
Hamiltonian on a scale of $D_N$, corresponding to the
temperature $T_N \sim D_N$.~\cite{Wilson75}

Due to the exponential form of the Boltzmann factors, the
reduced NRG basis set of ${\cal H}_N$ is sufficient for
an accurate calculation of thermodynamic quantities at
temperature $T_N$. This is no longer the case once the
system is driven out of equilibrium. Since the nonequilibrium
dynamics involves all energy scales exceeding $T_N$, and
in the absence of a general criterion as to which states
contribute to the dynamics, a complete basis set of the
Fock space ${\cal F}_N$ of ${\cal H}_N$ is required. In
the following we identify such a complete basis set of
${\cal F}_N$, composed of approximate eigenstates of
${\cal H}_N$.

\begin{figure}[btp]
\centering
\includegraphics[width=80mm]{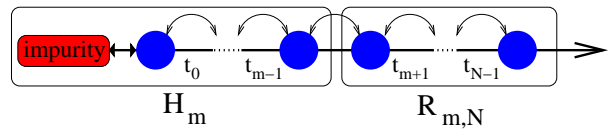}
\caption{The full Wilson chain of length $N$ is divided into
         a sub-chain of length $m$ and the ``environment''
         $R_{m,N}$. The Hamiltonian ${\cal H}_m$ can be
         viewed either as acting only on the sub-chain
         of length $m$, or as acting on the full chain of
         length $N$, but with the hopping matrix elements
         $t_m,\cdots,t_{N-1}$ all set to zero. The former
         picture is the traditional one. In this paper we
         adopt the latter point of view.} 
\label{fig:1}
\end{figure}

\subsubsection{Complete Basis Set }

There are two possible ways to envision the iterative
NRG solution of the $N$-site chain. In the traditional
picture one starts from a core cluster that consists of
the impurity degrees of freedom and the $n = 0$ site,
and enlarges the chain by one site at each NRG step.
Alternatively, one can view the NRG procedure as
starting from the full chain of length $N$, but with
the hopping matrix elements set to zero along the chain.
At each successive step another hopping matrix element is
switched on, until the full Hamiltonian ${\cal H}_N$ is
recovered. In this latter picture, to be adopted below,
the entire sequence of Hamiltonians ${\cal H}_m$ with
$m \leq N$ act on the Fock space of the $N$-site chain.
Accordingly, each NRG eigen-energy of ${\cal H}_m$ has
an extra degeneracy of $d^{(N-m)}$, where $d$ is the
number of distinct states at each site along the chain.
The extra degeneracy stems from the $N-m$ ``environment''
sites at the end of the chain, depicted in
Fig.~\ref{fig:1}.

Let us elaborate on the formal connection between the two
pictures presented above. Let $\{ |r; m\rangle \}$ label
the NRG eigenstates of ${\cal H}_m$ when acting on the
$m$-site chain, and let $E_{r}^{m}$ denote their corresponding
eigen-energies. Enumerating the different configurations of
site $i$ by $\{ \alpha_i \}$, each of the tensor-product states
$| r, \alpha_{m+1}, \cdots, \alpha_N \rangle$ with arbitrary
$\alpha_{m+1}, \cdots, \alpha_N$ is then a degenerate
eigenstate of ${\cal H}_m$ with energy $E_{r}^{m}$, when
acting on the full $N$-site chain. For brevity, we introduce
the shorthand notation $\ket{r,e;m}$, where the ``environment''
variable $e = \{\alpha_{m+1}, \cdots,\alpha_N\}$ encodes the
$N - m$ site labels $\alpha_{m+1}, \cdots, \alpha_N$. The
index $m$ is used in this notation to record where the chain
is partitioned into a ``subsystem'' and an ``environment''
(see Fig.~\ref{fig:1}).

In the conventional NRG picture, the eigenstates of
${\cal H}_{m+1}$ are constructed from the tensor-product
states $\{\ket{r,\alpha_{m+1};m} \}$, corresponding to
enlarging the chain by one extra site. A unitary
transformation $U$ relates the new eigenstates $\ket{r';m+1}$
of ${\cal H}_{m+1}$ to the basis $\{\ket{r,\alpha_{m+1};m} \}$: 
\begin{eqnarray}
\ket{r';m+1} &=& \sum_{r,\alpha_{m+1}} U_{r',r\alpha_{m+1}}
                 \ket{r,\alpha_{m+1};m} \; .
\label{eq:eigenstates-of-h-m+1}
\end{eqnarray}
An alternative notation is given by
\begin{eqnarray}
\ket{r';m+1} &=& \sum_{r,\alpha_{m+1}} P^m_{r',r}[\alpha_{m+1}]
                 \ket{r,\alpha_{m+1};m} \; ,
\label{eq:eigenstates-of-h-m+1-mps}
\end{eqnarray}
where the matrix elements $P_{r',r}[\alpha_{m+1}]$ are
identified with the corresponding matrix elements of the
unitary transformation $U$: $P^m_{r',r}[\alpha_{m+1}] =
U_{r',r\alpha_{m+1}}$. Successively applying the recursion
relation of Eq.~(\ref{eq:eigenstates-of-h-m+1-mps}), the
eigenstates of ${\cal H}_N$ can be formally viewed as
matrix-product states,~\cite{VmpsQIS} generated by
consecutive application of the matrices
$P^m_{r',r}[\alpha_{m+1}]$.

Note that the transition itself from the tensor-product
states $\{\ket{r,\alpha_{m+1};m} \}$ to the eigenstates
$\{\ket{r';m+1} \}$ involves no truncation of the Fock
space. However, since the dimension of the Fock space
grows as $d^N$, a complete basis set is not manageable
for any practical chain length of order $N \sim 100$. In
the equilibrium NRG, high-energy states are thus discarded
after each iteration, as these do not contribute to the
equilibrium density matrix. This latter statement is
guarantied by the hierarchy of scales along the Wilson
chain. It would not apply to any ordinary tight-binding
chain with constant hopping matrix elements.

Consider now the first iteration $m$ at which states are
discarded. In order to keep track of the complete basis
set of the $N$-site chain, we formally divide the
eigenstates $\ket{r,e;m}$ into two distinct subsets: the
discarded high-energy states $\{ \ket{l,e;m}_{dis} \}$ and
the retained low-energy states $\{ \ket{k,e;m}_{kp} \}$.
In the course of the NRG, only the latter states are used
to span the Hamiltonian $\H_{m+1}$ within the reduced
subspace $\{ \ket{k,\alpha_{m+1},e';m} \}$. Note, however,
that the sum of both subsets still constitutes a complete
basis set for the $N$-site chain.
Repeating this procedure at each subsequent iteration, we
recursively divide the retained subset into a discarded
part and a retained part. Then, the collection of {\em all
discarded} eigenstates $\ket{l,e;m}_{dis}$ together with
the eigenstates of the final NRG iteration $N$ combine
to form a complete basis set for the entire Fock space
${\cal F}_{N}$. Regarding all eigenstates of the final
NRG iteration as discarded, one can formally write the
Fock space of the $N$-site chain in the form
${\cal F}_{N} = {\rm span} \{|l,e;m \rangle_{dis}\}$.

We stress that {\em all states} are retained in the course
of this construction. Not a single state is eliminated. Since
$\{|l,e;m \rangle_{dis}\}$ constitutes a complete basis set
of ${\cal F}_{N}$, then the following completeness relation
obviously holds:
\begin{eqnarray}
\sum_{m = m_{\rm min}}^N \sum_{l,e}
          \ket{l,e;m}_{dis}\ _{dis}\bra{l,e;m} &=& 1 \; .
\label{equ:complete-basis}
\end{eqnarray}
Here the summation over $m$ starts from the first iteration
$m_{\rm min}$ at which a basis-set reduction is imposed.
Typical values of $m_{\rm min}$ are $4$-$5$ for a
spin-degenerate conduction bath. The summation indices
$l$ and $e$ implicitly depend on $m$. The identification
of this {\em complete basis set}, naturally generated by
the NRG, is one of the two key ingredients of our method.
All traces will be carried out below with respect to
this basis set. Hence, {\em the evaluation of time-dependent
expectation values involve no truncation error}. Note, however,
that we made no reference to a particular Hamiltonian ${\cal H}$
in constructing the basis set $\{ |l,e;m \rangle_{dis} \}$.
Below we shall make use of two distinct basis sets of this
form, one for the Hamiltonian ${\cal H}^f$ and another
for the unperturbed Hamiltonian ${\cal H}^i$.

Another useful identity to be used below pertains to the
subspace retained at iteration $m$, $\{ \ket{k,e;m}_{kp}\}$. To
this end, we note that the sum in Eq.~(\ref{equ:complete-basis})
can always be divided into two complementary parts $1_m^-$
and $1_m^+$: 
\begin{eqnarray}
1_m^- &=& \sum_{m'=m_{min}}^{m} \sum_{l',e'}
               \ket{l',e';m'}_{dis}\ _{dis}\bra{l',e';m'} \; ,
\label{eqn:1_m^-}
\\
1_m^+ &=& \sum_{m'= m+1}^N \sum_{l',e'}
               \ket{l',e';m'}_{dis}\ _{dis}\bra{l',e';m'} \; .
\label{eqn:1_m^+initial}
\end{eqnarray}
(For $m = N$ only $1_m^-$ exists.) The completeness relation
therefore becomes
\begin{eqnarray}
  1 &=& 1_m^- + 1_m^+ \; .
\label{eqn:completness}
\end{eqnarray}

What do the operators $1_m^-$ and $1_m^+$ represent? The
operator $1_m^-$ projects onto the subspace of all discarded
states up to and including iteration $m$. The operator
$1_m^+$ projects onto the complementary subspace retained
at iteration $m$. An alternative way of writing the latter
projection operator is in terms of the states retained
at iteration $m$, $\{\ket{k,e;m}_{kp}\}$. Explicitly,
$1_m^+$ is given by
\begin{equation}
1_m^+ = \sum_{k,e} \ket{k,e;m}_{kp} \, _{kp}\bra{k,e;m} \; .
\label{eqn:1_m^+}
\end{equation}

Equations~(\ref{eqn:1_m^+initial}) and
(\ref{eqn:1_m^+}) provide us with an important
connection between summations over retained and discarded
states. These equivalent representations of $1_m^+$ will
be frequently used later on.

\subsubsection{Time-evolution of local expectation values}

We are now in position to formally evaluate the time-dependent
expectation value of any local operator $\hat O$ at time $t\ge 0$.
As indicated in Eq.~(\ref{eqn:time-evolution-general}), we need
to explicitly carry out the trace over $\hat{\rho}(t) \hat{O}$.
This is most conveniently done using the basis set
$\{ |l,e;m \rangle_{dis} \}$ introduced in the previous section:
\begin{eqnarray}
O(t) &=& \!\!\sum_{m = m_{\rm min}}^{N} \sum_{l,e}
              \, _{dis}\langle l,e;m| \hat \rho(t) \, \hat{O}
              |l,e;m\rangle_{dis} \; .
\label{O-of-t}
\end{eqnarray}
Inserting Eq.~(\ref{eqn:completness}) in between the operators
$\hat\rho(t)$ and $\hat O$ in Eq.~(\ref{O-of-t}) yields
\begin{equation}
O(t) = \!\! \sum_{m = m_{\rm min}}^{N} \sum_{l,e}
            \, _{dis}\langle l,e;m| \hat \rho(t) \!
            \left(1_m^- +1_m^+\right)\!
            \hat{O} |l,e;m\rangle_{dis} 
\; .
\label{O-of-t-ii}  
\end{equation}
Using Eqs.~(\ref{eqn:1_m^-}) and (\ref{eqn:1_m^+})
for $1_m^-$ and $1_m^+$, respectively, we obtain
$\hat O (t) = O_-(t) + O_0(t) + O_+(t)$, where
\begin{eqnarray}
O_- (t) \! &=& \!\!\!\!
\sum_{m > m_{\rm min}\,}^{N}
\sum_{m'=m_{\rm min}}^{m - 1}
\sum_{l,l'}
\sum_{e,e'} 
        \, _{dis}\langle l',e';m'| {\hat O} |l,e;m \rangle_{dis}
\nonumber \\
&& \times \, _{dis}\langle l,e;m| \hat \rho(t)
          |l',e';m'\rangle_{dis}
\end{eqnarray}
and
\begin{eqnarray}
O_0 (t) &=& \!\! \sum_{m = m_{\rm min}}^{N}
            \sum_{l,l'} \sum_{e,e'}
            \, _{dis}\langle l,e;m| \hat \rho(t)
                            |l',e';m\rangle_{dis}
\nonumber \\
&& \times \, _{dis}\langle l',e';m| {\hat O}
                            |l,e;m \rangle_{dis}
\label{eqn:O_0}
\end{eqnarray}
stem for the projection operator $1_m^-$, and
\begin{eqnarray}
O_+(t) &=& \!\! \sum_{m = m_{\rm min}}^{N - 1}
                \sum_{l,k} \sum_{e,e'}
                \, _{dis} \langle l,e;m| \hat \rho(t)
                         |k,e';m\rangle_{kp}
\nonumber \\
&& \times \, _{kp}\langle k,e';m| {\hat O}
                          |l,e;m \rangle_{dis}
\label{eqn:O_plus}
\end{eqnarray}
originates from $1_m^+$.

Whereas the terms $O_{0} (t)$ and $O_{+}(t)$ involve the
summation over a single iteration variable $m$, $O_{-}(t)$
contains two such summations: one over $m$ and another
over $m'$. Rearranging the two sums according to 
\begin{equation}
\sum_{m = m_{\rm min} + 1\,}^{N}
     \sum_{m'=m_{\rm min}}^{m - 1} \to
\sum_{m'=m_{\rm min}\,}^{N - 1} \sum_{m = m'+1}^{N} \; ,
\end{equation}
utilizing the identity
\begin{equation}
\sum_{m = m'+1}^{N} \sum_{l, e} |l,e;m \rangle_{dis}
     \, _{dis}\langle l,e;m| = 1_{m'}^+ \; ,
\end{equation}
and replacing $1_{m'}^+$ by the representation of
Eq.~(\ref{eqn:1_m^+}) we obtain
\begin{eqnarray}
O_-(t) &=&  
\sum_{m' = m_{\rm min}}^{N - 1} \sum_{k, l'}\sum_{e, e'}
     \, _{kp}\langle k,e;m'| {\hat \rho}(t)
     |l',e';m' \rangle_{dis}
\nonumber \\
&& \times \, _{dis}\langle l',e';m'| \hat{O}
     |k,e;m'\rangle_{kp} \; .
\label{eqn:O_minus}
\end{eqnarray}
Combining Eqs.~(\ref{eqn:O_0}), (\ref{eqn:O_plus}) and
(\ref{eqn:O_minus}) then yields
\begin{eqnarray}
O(t) &=& \!\!\sum_{m = m_{\rm min}}^{N}
             \sum_{r,s}^{trun} \sum_{e,e'} \;
                  \langle s,e;m| \hat \rho(t) |r,e';m\rangle
\nonumber \\
&& \times \langle r,e';m| {\hat O} |s,e;m \rangle \; .
\label{O-of-t-II}
\end{eqnarray}
Here the restricted sum $\sum^{trun}_{r,s}$ implies that
at least one of the states $r$ and $s$ is {\em discarded}
at iteration $m$. Terms where both states are retained
contribute to the sum at some later iteration $m' > m$.

So far we made no assumption about the operator $\hat{O}$,
nor have we committed ourselves to a particular Hamiltonian
in writing the basis set $\{ |l,e;m \rangle_{dis} \}$.
In the following we restrict ourselves to local operators
$\hat{O}$. By local we mean that the operator acts on
degrees of freedom that reside either on the impurity or
on close-by sites $\bar m$ for which all states are still
available (i.e., $\bar m \le m_{min}$). This restriction
is rather weak, since all operators in the vicinity of the
impurity fulfill this requirement. The matrix elements of
a local operator $\hat{O}$ are diagonal in and independent
of the environment degrees of freedom: 
\begin{equation}
\langle r,e;m| {\hat O} |s,e';m \rangle = 
        \delta_{e,e'} O_{r,s}^m \; .
\end{equation}
This allows us to write Eq.~(\ref{O-of-t-II}) in the form
\begin{equation}
O(t) = \!\!\sum_{m = m_{\rm min}}^{N}
           \sum_{r,s}^{trun} \sum_{e} \; O_{r,s}^m
           \langle s,e;m| \hat \rho(t) |r,e;m\rangle \; .
\label{O-of-t-III}
\end{equation}

As for the states $\{ |l,e;m \rangle_{dis} \}$, unless
stated otherwise we work hereafter in the NRG basis set
generated for the perturbed Hamiltonian ${\cal H}^f$.
This choice is motivated by the time dependence of
$\hat{\rho}(t)$, which is governed by ${\cal H}^f$.
Indeed, using Eq.~(\ref{rho-of-t}) and resorting
to the standard NRG approximation~\cite{Wilson75}
$H_N^f \ket{k,e;m} \approx  E_k^m \ket{k,e;m}$ one
has
\begin{equation}
\langle s,e;m| \hat \rho(t) |r,e;m\rangle \approx
        e^{i(E_r^m - E_s^m)t}
        \langle s,e;m| \hat{\rho}_0 |r,e;m\rangle \; .
\label{eqn:standard-NRG-approx}
\end{equation}
Upon inserting Eq.~(\ref{eqn:standard-NRG-approx})
into Eq.~(\ref{O-of-t-III}), the environment variable $e$
enters only through the matrix element of $\hat{\rho}_0$.
Introducing the reduced density
matrix~\cite{AndersSchiller2005} 
\begin{equation}
\rho^{\rm red}_{s,r}(m) = \sum_{e}
          \langle s,e;m|\hat{\rho}_0 |r,e;m \rangle \; ,
\label{eqn:reduced-dm-def}
\end{equation}
one arrives at the final result for the time evolution
of $O(t)$ at $t \ge 0$:~\cite{AndersSchiller2005}
\begin{equation}
O(t) = \sum_{m = m_{\rm min}}^{N} \sum_{r,s}^{trun} \;
       e^{i(E_{r}^m -E_{s}^m)t} O_{r,s}^m
       \rho^{\rm red}_{s,r}(m) \; .
\label{eqn:time-evolution}   
\end{equation}

Several comments should be made about
Eq.~(\ref{eqn:time-evolution}). First, no assumption was
made about the initial density operator $\hat{\rho}_0$.
It can either project onto a particular state or represent
the equilibrium density operator for ${\cal H}^i$. It can
even stand for a density operator that has already evolved
in time subject to some previous dynamics. Second, all
states of the finite-size Fock space are retained in
Eq.~(\ref{eqn:time-evolution}) and all energy scales are
explicitly taken into account. No basis set reduction
is imposed. This should be contrasted with recent
time-dependent extensions of the density-matrix
renormalization group.~\cite{Marston2002,
DaleyKollathSchollwoeckVidal2004, WhiteEeiguin2004,
GobertTdDMRG2004,SchollwoeckDMRG2005} In the so-called
TD-DMRG, the ground state is evolved in time using a
Trotter decomposition of the time-evolution operator.
This introduces an accumulated error that grows linearly
in time. Here $O(t)$ is evaluated independently for each
time $t \ge 0$, and is thus free of accumulated errors.

This does not mean that the Eq.~(\ref{eqn:time-evolution})
is error-free. Two approximations enter the TD-NRG:
(i) a discretized finite-size representation of the
continuous bath, and (ii) the conventional NRG approximation
$H_N^f \ket{k,e;m} \approx E_k^m \ket{k,e;m}$. Indeed, the
latter approximation is the only one invoked in describing
the time evolution of expectation values on the $N$-site
chain. In particular, Eq.~(\ref{eqn:time-evolution}) is
exact for $t \to 0^+$. As shown by Wilson,~\cite{Wilson75}
the associated error in thermodynamic quantities is
perturbative and small, a consequence of the separation
of scales along the Wilson chain. To assess the accuracy
of this approximation in the present context we have
compared Eq.~(\ref{eqn:time-evolution}) to the exact
time evolution on sufficiently short chains where all
eigenstates of ${\cal H}^f$ can be computed and stored
in memory. To facilitate a meaningful comparison, only
a deliberately small number of the states were kept in
the course of the NRG iterations. A surprisingly good
agreement was found between the exact results and the
TD-NRG up to very long times, establishing the accuracy
of the accepted NRG approximation in the present context.

A more significant source of error is due to the
discretized finite-size representation of the continuous
bath. There are two aspects to this approximation. Due
to the limited energy resolution at low energies,
Eq.~(\ref{eqn:time-evolution}) generally looses accuracy
for $t \gg 1/D_N$. Since the chain length $N$ is chosen
such that $D_N \sim T$, one expects then a lose of accuracy
for $t \gg 1/T$. It should be noted, however, that the
NRG can access arbitrarily low temperatures, implying that
arbitrarily long time scales can be reached for $T \to 0$.
We demonstrate this important point later on. 
At the same time, a continuous spectrum is expected to be  vital for
relaxation  to the exact thermodynamic expectation value with respect
to $\H^f$. This constitutes a fundamental obstacle for any
solution, however accurate, of a finite-size system.
As detailed in subsection~\ref{sec:z-trick},
we can largely overcome this obstacle by (i) averaging over
different realizations of the Wilson chain using Oliveira's
$z$-trick,~\cite{YoshidaWithakerOliveira1990} and (ii)
resorting to a Lorentzian broadening of the NRG levels.

\subsection{Calculation of the reduced density matrix}
\label{sec:reduced-density-matrix}

So far, our discussion applied to a general local operator
$\hat{O}$ and to an arbitrary initial density operator
$\hat{\rho}_0$. The time evolution of $O(t)$ was shown
to be determined by a sequence of reduced density matrices
$\rho^{\rm red}$, in which the environment degrees of
freedom are traced out iteratively. While formally
applicable to any $\hat{\rho}_0$, practical calculations
may prove more restrictive. Here we address the
calculation of the reduced density matrix.

Typically, $\hat{\rho}_0$ has a simple representation
with respect to the eigenstates of the unperturbed
Hamiltonian ${\cal H}^i$. For example, starting from
thermal equilibrium $\hat{\rho}_0$ is equal to
$e^{-\beta{\cal H}^i}/Z_i$, where $Z_i = {\rm Tr}
\{ e^{-\beta{\cal H}^i}\}$ is the unperturbed partition
function. Thus, the reduced density matrix is most
conveniently expressed with respect to the NRG
eigenstates of the unperturbed Hamiltonian, i.e.,
when the states $\{|l,e;m\rangle\}$ in
Eq.~(\ref{eqn:reduced-dm-def}) relate to ${\cal H}^{i}$.
Note, however, that the reduced density matrix which
enters Eq.~(\ref{eqn:time-evolution}) was explicitly
constructed for the NRG eigenstates of the full
Hamiltonian ${\cal H}^{f}$. In general, there is no
simple relation between the two reduced density matrices,
as they involve traces over different environments.
Still, one can convert between the two entities in the
generic case where the system is evolved in time from
thermal equilibrium. Below we explain in detail how the
reduced density matrix of Eq.~(\ref{eqn:reduced-dm-def})
is calculated for this generic case.

\subsubsection{General considerations}
\label{sec:general-considerations}

Let us assume for the time being that the reduced density
matrix is know with respect to the NRG eigenstates
of the unperturbed Hamiltonian ${\cal H}^i$. Namely,
when the states $\{|l,e;m\rangle\}$ that enter
Eq.~(\ref{eqn:reduced-dm-def}) correspond to ${\cal H}^{i}$.
Our goal is to compute the reduced density matrix with
respect to the NRG eigenstates of the full Hamiltonian
${\cal H}^{f}$. Technically, this requires working
with two distinct sets of NRG states --- one for
${\cal H}^{i}$ and the other for ${\cal H}^{f}$. To
simplify the notation as much as possible, we distinguish
the two sets of states by their labels. Throughout this
subsection we designate the NRG states pertaining to
the unperturbed Hamiltonian ${\cal H}^{i}$ by indices
carrying the subscript $i$. For example, the state
$|l_i,e_i;m\rangle$. NRG states corresponding to
${\cal H}^{f}$ will be labeled, as before, with plain
indices having no additional subscripts as in
$|l,e;m\rangle$. We stress that $e_i$ and $e$ label
the same environment degrees of freedom, despite the
difference in appearance. Similarly, we reserve the
notation
$\rho^{\rm red}_{s, r}(m)$ for the reduced density
matrix as defined in Eq.~(\ref{eqn:reduced-dm-def})
with respect to the states of the full Hamiltonian
${\cal H}^{f}$. The reduced density matrix with
respect to the states of the unperturbed Hamiltonian
${\cal H}^{i}$ is denoted (and defined) by
\begin{equation}
\rho_{s_i,r_i}^{{\rm red}, 0}(m) = \sum_{e_i} \,
     \langle s_i,e_i;m| \hat{\rho}_0 |r_i,e_i;m\rangle \; .
\label{eqn:reduced-equilibrium}
\end{equation}

As shown in the previous subsection, one can write the
completeness relation $1 = 1_m^{-} + 1_m^{+}$ for any
Hamiltonian ${\cal H}$. The projection operators $1_m^{-}$
and $1_m^{+}$ are simply written in this case using the
NRG states of ${\cal H}$ [see Eqs.~(\ref{eqn:1_m^-}) and
(\ref{eqn:1_m^+})]. Below we apply this completeness
relation for the initial Hamiltonian ${\cal H}^{i}$,
with one slight modification. Shifting the $m' = m$ term
from Eq.~(\ref{eqn:1_m^-}) to Eq.~(\ref{eqn:1_m^+}) one
has the identity
\begin{equation}
1 = {\cal I}_m^{-} + {\cal I}_m^{+} \; ,
\label{eqn:completenes-via-cal-I}
\end{equation}
where
\begin{equation}
{\cal I}_m^- = \sum_{m'=m_\textrm{min}}^{m-1}
               \sum_{l'_i,e'_i}
                    |l'_i,e'_i;m'\rangle_{dis}\
                    _{dis}\langle l'_i,e'_i;m'|
\label{eqn:cal-I_m^-}
\end{equation}
and
\begin{equation}
{\cal I}_m^+ = \sum_{k_i,e_i}
               |k_i,e_i;m\rangle \langle k_i,e_i;m| \; .
\label{eqn:cal-I_m^+}
\end{equation}
Here the index $k_i$ in Eq.~(\ref{eqn:cal-I_m^+}) runs
over all NRG eigenstates of iteration $m$, whether
discarded or retained. This shift of notation from $1_m$ to 
${\cal I}_m$ is of purely practical nature as it allows us to sum
freely over all states $k_i$ of any given iteration $m$. 
The symbol ${\cal I}$ is used to emphasize the different complete
basis sets of $\H^f$ and $\H^i$.

We are now in position to address the reduced density
matrix $\rho^{\rm red}_{s, r}(m)$. Starting from
Eq.~(\ref{eqn:reduced-dm-def}), we make use of the
completeness relation of
Eq.~(\ref{eqn:completenes-via-cal-I}) in order to
replace $\hat{\rho}_0$ with
\begin{equation}
\left( {\cal I}_m^{-} + {\cal I}_m^{+} \right)
       \hat{\rho}_0
\left( {\cal I}_m^{-} + {\cal I}_m^{+} \right) .
\end{equation}
Equation~(\ref{eqn:reduced-dm-def}) is decomposed in
this fashion into four contributions:
\begin{equation}
\rho^{\rm red}_{s,r}(m) =
     \rho^{++}_{s,r}(m) + \rho^{+-}_{s,r}(m) +
     \rho^{-+}_{s,r}(m) + \rho^{--}_{s,r}(m) \; ,
\end{equation}
where $\rho^{p p'}_{s,r}(m)$ equals
\begin{equation}
\rho^{p p'}_{s,r}(m) = \sum_{e}
     \langle s,e;m| {\cal I}_m^{p} \hat{\rho}_0
             {\cal I}_m^{p'} |r,e;m \rangle
\label{eqn:expansion-of-rho-red}
\end{equation}
($p, p' = \pm$). Of these four components, only
$\rho^{++}_{s,r}(m)$ can be directly related to
the ``unperturbed'' reduced density matrix of
Eq.~(\ref{eqn:reduced-equilibrium}). To see this,
we insert Eq.~(\ref{eqn:cal-I_m^+}) into
Eq.~(\ref{eqn:expansion-of-rho-red}) to obtain
\begin{eqnarray}
\!\!\!\!\!\!\!\!
\rho^{++}_{s,r}(m) \!\! &=& \!\!
          \sum_{e} \sum_{k_i, k'_i}
          \sum_{e_i,e'_i}
          \langle s,e;m | k'_i,e'_i;m \rangle
\nonumber \\
&& \!\! \times \langle k'_i,e'_i;m | \hat{\rho}_0
               | k_i,e_i;m \rangle
               \langle k_i,e_i;m | r,e;m \rangle .
\end{eqnarray}
This expression features the overlap matrix elements
$\langle k_i, e_i;m | r,e;m \rangle$, which are diagonal
in and independent of the environment degrees of freedom: 
\begin{eqnarray}
\langle k_i, e_i;m | r,e;m \rangle =
           \delta_{e, e_i} S_{k_i,r}(m) \; .
\label{eq:S_kr-def}
\end{eqnarray}
The reduced matrix $S(m)$ records the overlap matrix
elements between the NRG eigenstates of ${\cal H}_m^{i}$
and ${\cal H}_m^{f}$. It is conveniently computed in
the course of the NRG iterations as described in
Appendix~\ref{app:overlap-matrix}. With the matrix
$S(m)$ at hand, $\rho^{++}_{s,r}(m)$ is obtained by
a simple rotation of the ``unperturbed'' reduced
density matrix into the new basis:
\begin{eqnarray}
\rho^{++}_{s,r}(m) = \sum_{k_i,k'_i}
     S^{\ast}_{k'_i,s}(m) \rho_{k'_i,k_i}^{{\rm red},0}(m)
     S_{k_i,r}(m) \; .
\label{eqn:rho-pp-m}
\end{eqnarray}
Note, however, that this transformation is not unitary due to basis
set reduction  at each iteration. Hence, $\rho^{++}_{s,r}(m)$
follows directly from the knowledge of $\rho^{{\rm red},0}_{s,r}(m)$.

In contrast to $\rho^{++}_{s,r}(m)$, the remaining
components $\rho^{+-}_{s,r}(m)$, $\rho^{-+}_{s,r}(m)$
and $\rho^{--}_{s,r}(m)$ cannot be related in any
simple way to the ``unperturbed'' density matrix
$\rho^{{\rm red},0}(m)$. This is readily seen by
inserting Eq.~(\ref{eqn:cal-I_m^-}) into
Eq.~(\ref{eqn:expansion-of-rho-red}), which gives
rise to overlap matrix elements of the form
$_{dis}\langle k_i, e_i;m' | r,e;m \rangle$ with
$m' < m$. The latter matrix elements generally depend
on both $e_i$ and $e$, which prevents from collapsing
the trace over $e_i$ to the form that appears in
Eq.~(\ref{eqn:reduced-equilibrium}). At present we have
no feasible way to carry out such a trace.

Physically, the terms $\rho^{+-}(m)$, $\rho^{-+}(m)$
and $\rho^{--}(m)$ encode the way in which high- and
low-energy eigenstates of ${\cal H}^i$ are coupled
within $\hat{\rho}_0$. For these terms to be important,
$\hat{\rho}_0$ must contain a significant contribution
from high-energy states, which means starting from a
state well removed from thermal equilibrium. In the
 following we restrict attention to the case where
$\hat{\rho}_0$ corresponds to thermal equilibrium, or
more generally to the case where
$\hat{\rho}_0 = {\cal I}^+_N \hat{\rho}_0 {\cal I}^+_N$.

\begin{figure}[tbp]
(a) \includegraphics[height=20mm]{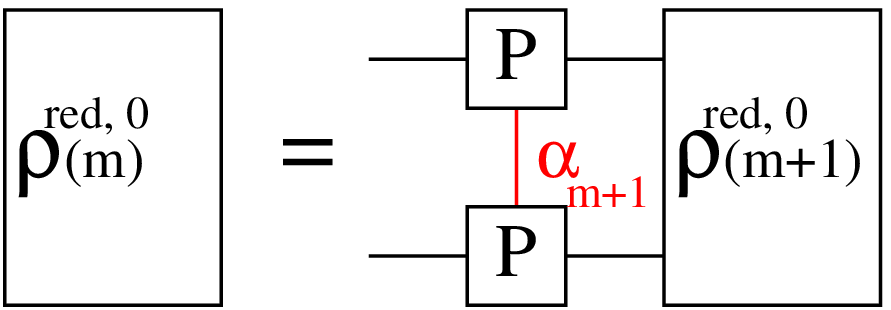}
    \hfill \\[6mm]
(b) \includegraphics[height=15mm]{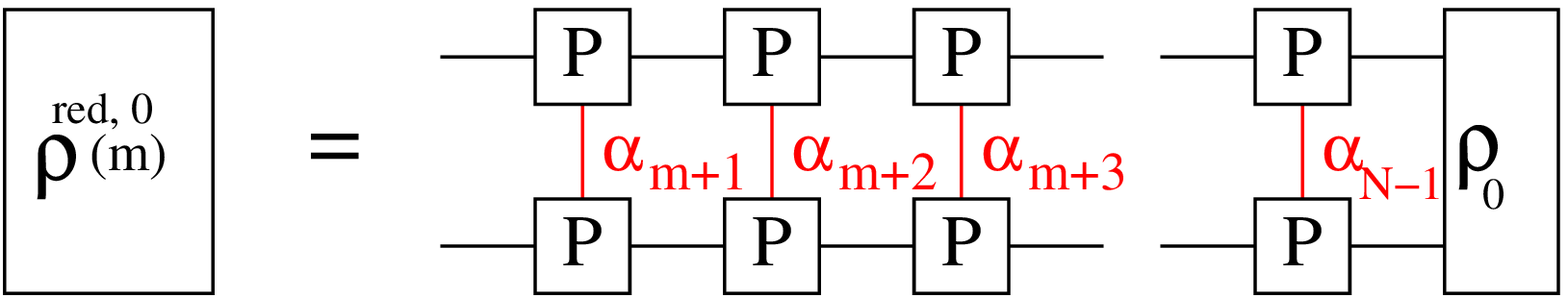}
\caption{(a) Diagrammatic representation of the recursion
    relation of Eq.~(\ref{eqn:redu-matrix-recursion}) for
    $\rho_{s_i,r_i}^{{\rm red},0}(m)$. Each box stands for
    a matrix element $P_{l'_i,l_i}[\alpha_{m+1}]$ (upper
    row) or its complex conjugate (lower row). The state
    labels $l_i$ and $l'_i$ are plotted horizontally, with
    $l_i$ to the left and $l'_i$ to the right. The state
    label $\alpha_{m+1}$ for the $m+1$ site is plotted
    vertically. A connected line indicates summation over
    the corresponding index. The reduced density matrix
    $\rho_{s_i,r_i}^{{\rm red},0}(m)$ has two indices
    $s_i$ and $r_i$, carried by the external legs on the
    left. Iterating (a) up to $m' = N$ relates the reduced
    density matrix $\rho_{s_i,r_i}^{{\rm red},0}(m)$ to
    the equilibrium NRG density matrix
    $\langle k'_i;N| \hat{\rho}_0 | k_i,N \rangle$. This
    process is visualized diagrammatically in (b). The
    sequence of connected vertical lines amounts to
    tracing over the environment degrees of freedom
    $\{\alpha_i\}_{i=m+1}^{N}$.} 
\label{fig:mps-rho-red}
\end{figure}

\subsubsection{Starting from thermal equilibrium}

Upon starting from thermal equilibrium, $\hat{\rho}_0$
takes the standard form $e^{-\beta{\cal H}^i}/Z_i$,
where $Z_i = {\rm Tr} \{ e^{-\beta{\cal H}^i}\}$ is
the unperturbed partition function. In terms of the
appropriate NRG basis, it has the spectral
representation~\cite{Wilson75} 
\begin{equation}
\hat{\rho}_0 = \frac{1}{Z_i}
           \sum_{l_i} e^{-\beta E^N_{l_i}}
                      | l_i;N \rangle \langle l_i;N | \; ,
\label{eqn:spectral-rho-0}
\end{equation}
\begin{equation}
Z_i = \sum_{l_i} e^{-\beta E^N_{l_i}} \; .
\end{equation}
These expressions neglect all states discarded at
iterations $m < N$, exploiting the fact that the
discarded states have only an exponentially small
contribution to $\hat{\rho}_0$ at the temperature
$T_N$.~\cite{Wilson75} This NRG approximation
usually works very well, and can be systematically
improved by increasing the number of states retained
at the conclusion of each NRG iteration. We denote
the latter number of states by $N_s$.

By construction, Eq.~(\ref{eqn:spectral-rho-0}) obeys
the operator identity
$\hat{\rho}_0 = {\cal I}^+_N \hat{\rho}_0 {\cal I}^+_N$.
This guarantees that $\rho^{+-}_{s,r}(m)$,
$\rho^{-+}_{s,r}(m)$ and $\rho^{--}_{s,r}(m)$
identically vanish for all $m \le N$. Hence
$\rho^{\rm red}_{s,r}(m) = \rho^{++}_{s,r}(m)$ is
fully determined by $\rho^{{\rm red}, 0}_{s_i,r_i}(m)$
according to Eq.~(\ref{eqn:rho-pp-m}). We now elaborate
how $\rho^{{\rm red}, 0}(m)$ is computed recursively
from $\rho^{{\rm red}, 0}(m+1)$. Together with the
``initial'' condition
\begin{equation}
\rho^{{\rm red}, 0}_{s_i,r_i}(N) = \delta_{s_i, r_i}
     \frac{1}{Z_i} e^{-\beta E^N_{s_i}} \; ,
\end{equation}
this provides us with $\rho^{{\rm red}, 0}(m)$ for
all $m \leq N$.

Consider an arbitrary $m < N$. To execute the sum over
$e_i$ in Eq.~(\ref{eqn:reduced-equilibrium}), we set
$e_i = (\alpha_{m+1}, e'_i)$, where $e'_i$ encodes the
$N - m + 1$ site labels $\alpha_{m+2}, \cdots, \alpha_N$.
In other words, $e'_i$ is a state variable for the
environment $R_{m+1, N}$, see Fig.~\ref{fig:1}.
Substituting ${\cal I}^+_{m+1} \hat{\rho}_0
{\cal I}^+_{m+1}$ in for $\hat{\rho}_0$, and using
the overlap matrix elements
\begin{equation}
\langle s_i, e_i; m| k'_i, e'_i; m+1 \rangle =
    P_{k'_i; s_i}[\alpha_{m+1}]
\end{equation}
[see Eq.~(\ref{eq:eigenstates-of-h-m+1-mps})],
Eq.~(\ref{eqn:reduced-equilibrium}) is expressed as
\begin{eqnarray}
\rho_{s_i,r_i}^{{\rm red},0}(m) &=& 
     \sum_{\alpha_{m+1}} \sum_{k_i, k'_i}^{\rm ret}
          P_{k'_i;s_i}[\alpha_{m+1}]
          P_{k_i;r_i}^{\ast}[\alpha_{m+1}]
\nonumber \\
&& \times
     \rho_{k'_i,k_i}^{{\rm red},0}(m+1) \; ,
\label{eqn:redu-matrix-recursion}
\end{eqnarray}
where the sum over $k_i$ and $k'_i$ is restricted to the
states retained at iteration $m+1$. (For $m = N-1$, the sum
runs over all the states of the final NRG iteration). In
the standard case of a real Hamiltonian ${\cal H}^i$, the
matrix elements $P_{l'_i;l_i}[\alpha_{m+1}]$ are likewise
real. Under these circumstances one can omit the complex
conjugate from $P_{k_i;r_i}^{\ast}[\alpha_{m+1}]$. It is
also easy to verify that $\rho_{s_i,r_i}^{{\rm red},0}(m)$
vanishes if at least one of the states $s_i$ and $r_i$
is discarded at iteration $m$. This is a simple consequence of the
orthogonality of our basis-set states $\ket{l_i,e_i;m}$ and the
spectral representation (\ref{eqn:spectral-rho-0}), i.e.\ 
$\bra{l_i,e_i;m}\hat \rho_0 \ket{l'_i,e'_i;m}=0$, for all $m <N$.
The recursion relation can also be visualized
diagrammatically depicted in Fig.\ \ref{fig:mps-rho-red} as brought to
our attention by Weichselbaum.  

Note that the recursion relation of
Eq.~(\ref{eqn:redu-matrix-recursion}) is not confined to
thermal equilibrium. It relied solely on the fact that
$\hat{\rho}_0 = {\cal I}^+_N \hat{\rho}_0 {\cal I}^+_N$,
which guaranteed that $\hat{\rho}_0 = {\cal I}^+_{m+1}
\hat{\rho}_0 {\cal I}^+_{m+1}$ for all $m < N$. No
additional restriction was made to an equilibrium
distribution. Nevertheless, one is typically interested
in the case where the system starts from thermal
equilibrium. The ``unperturbed'' reduced density
matrix  $\rho^{{\rm red}, 0}(m)$ coincides then with
the one originally introduced by Hofstetter for the
purpose of calculating equilibrium spectral functions
within the NRG.~\cite{Hofstetter2000}

\subsection{TD-NRG algorithm}
\label{sec:td-nrg-algo}

After the exposition of the different components of the
TD-NRG approach, we now turn to the integrated algorithm.
To implement the TD-NRG at temperature $T$, one first
selects a chain length $N$ such that $T \approx T_N$.
Two simultaneous NRG runs are then performed, one for
${\cal H}^i$ and another for ${\cal H}^f$. All NRG
eigen-energies of ${\cal H}^i_m$ and ${\cal H}^f_m$
are stored up to the final iteration $N$. At each
iteration $m$, the overlap matrix $S_{r_i,r}(m)$ of
Eq.~(\ref{eq:S_kr-def}) is calculated between all
eigenstates $|r_i, m\rangle$ and $|r, m\rangle$ of
${\cal H}_m^i$ and ${\cal H}_m^f$, respectively.
Details of the calculation are elaborated in
Appendix~\ref{app:overlap-matrix}. This information,
as well as the product matrices $P_{l',l}[\alpha_m]$
for the initial and final Hamiltonian, are stored on the hard drive.
After the final NRG run, the equilibrium density
matrix is calculated from Eq.~(\ref{eqn:spectral-rho-0})
using the eigenstates and eigen-energies of last NRG
iteration for ${\cal H}^i_N$. 

At the conclusion of these steps, the TD-NRG proceeds
by backward iterations starting from $m = N$. For each
backward iteration from $m$ to $m-1$, the following
three steps are performed:
\begin{enumerate}

\item
   The ``unperturbed'' density matrix
   $\rho^{{\rm red}, 0}(m-1)$ is calculated from
   $\rho^{{\rm red}, 0}(m)$ using the product matrices
   $P_{l'_i,l_i}[\alpha_m]$ for the initial Hamiltonian
   ${\cal H}^i_m$, in combination with the recursion
   relation of Eq.~(\ref{eqn:redu-matrix-recursion}).

\item
   The reduced density matrix $\rho^{\rm red}_{s,r}(m-1)$
   is computed by rotating $\rho^{{\rm red}, 0}(m-1)$ to
   the basis of the final Hamiltonian using the overlap
   matrix $S(m-1)$ and Eq.~(\ref{eqn:rho-pp-m}).

\item
   Using Eq.~(\ref{eqn:time-evolution}), the contribution
   of iteration $m$ to $O(t_j)$ is evaluated simultaneously
   for all times $t_j$ of interest. Here the only matrix
   elements of $\rho_{s, r}^{\rm red}(m)$ and $O_{r, s}^m$
   to contribute are those where at least one of the states
   $s$ and $r$ is discarded at iteration $m$. At the
   conclusion of this step, $\rho^{\rm red}(m)$ is deleted
   to reduce the memory load.
\end{enumerate}
These steps are repeated until $m = m_{\rm min}$ is
reached, below which no state has been discarded.

It is easy to see that if ${\cal H}^f = {\cal H}^i$,
i.e., the Hamiltonian is left unchanged, then $O(t)$
so obtained coincides with the thermodynamic average
of $\hat{O}$ for all $t$. Indeed, the overlap matrix
$S(m)$ is diagonal in this case, which means that
$\rho^{\rm red}(m)$ and $\rho^{{\rm red}, 0}(m)$ are
the same. Consequently, $\rho^{\rm red}_{s, r}(m)$
has nonzero matrix elements only if the states $r$
and $s$ are both retained, leaving only the iteration
$m = N$ to contribute to Eq.~(\ref{eqn:time-evolution}).
Since $\rho^{\rm red}(N)$ coincides with the equilibrium
density matrix which is diagonal in energy, one recovers
the thermodynamic average of $\hat{O}$ independent of
$t$.

A word is in order at this point about the nonequilibrium
NRG approach of Costi,~\cite{Costi97} and its relation
to the TD-NRG. Costi had focused on the calculation of
nonequilibrium spectral functions. In contrast to the
TD-NRG, he settled with a single NRG shell $m$ to evaluate
the spectral function at frequency $\omega \sim D_m$.
In practice this meant replacing the reduced density
matrix $\rho^{{\rm red}, 0}_{s_i, r_i}(m)$ with the full
equilibrium density matrix at iteration $m$. After
rotating the latter according to Eq.~(\ref{eqn:rho-pp-m}),
only a single Wilson shell contributes to a given $\w$ of the spectral
representation of $O(t)$. Being well aware
that a full multiple-shell evaluation is ultimately
required for the correct description of nonequilibrium
dynamics,~\cite{Costi97} Costi carefully applied the
single-shell approach only to intermediate frequencies.
This restricted his ability to track the real-time
dependence of $O(t)$.
                                                                                
In the TD-NRG we overcame these difficulties by
identifying an appropriate basis set for the Hilbert
space of the $N$-site chain, and by the resummation
procedure that has led to Eq.~(\ref{eqn:time-evolution}).
The significance of these steps is best reflected in the
fact that $O(t \to 0^+)$ of Eq.~(\ref{eqn:time-evolution})
exactly coincides for any local operator $\hat{O}$
with its NRG thermodynamic average with respect to
${\cal H}^i_N$. While physically clear, this statement
is highly nontrivial from the standpoint of the TD-NRG,
as {\em all energy scales} contribute to the summation
of Eq.~(\ref{eqn:time-evolution}).
Moreover, the limit $O(t\to \infty)$ stems from the degenerate
terms with $E_r^m = E_s^m$ from {\em all} iterations $m$ in
Eq.~(\ref{eqn:time-evolution}). In frequency domain, these terms give
rise to an addition $\delta(\w)$ contribution that comes out naturally
in the TD-NRG, but is absent in Costi's approach. The latter is
of crucial importance for obtaining the correct asymptotic
long-time behavior of operators with non-vanishing expectation values.

\subsection{Towards restoring the continuum limit}
\label{sec:z-trick}

So far, we focused on accurately calculating the time
evolution of any local observable $\hat{O}$ on a
finite-length chain. While this might be a reasonable
representation of mesoscopic systems characterized by
a finite level spacing, our approach is geared toward
the description of a continuous bath. Relaxation in
such a macroscopically large system cannot be fully
described by a finite Wilson chain. Indeed, it was
already pointed out in the context of the equilibrium
NRG~\cite{YoshidaWithakerOliveira1990} that certain
thermodynamical properties undergo unphysical oscillations
as a result of the discretization of the continuous
bath. To circumvent this problem, Oliveira and
coworkers~\cite{YoshidaWithakerOliveira1990} introduced
a $z$-dependent logarithmic discretization of the
continuous bath according to
$[1,\Lambda^{-z},\Lambda^{-z-1},\cdots,\Lambda^{-z-n-1},\cdots]$.
Wilson's original discretization corresponds in this
notion to $z=1$. As shown by Oliveira {\em et al}., the
unphysical oscillations can be removed by integrating
expectation values with respect to $0 < z \leq 1$. This
has the effect of mimicking a continuous bath using a
single adjustable parameter.

\begin{figure}[tbp]
  \centering
  \includegraphics[width=85mm]{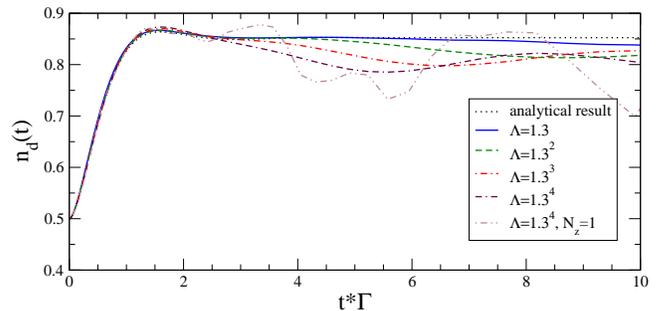}
  \caption{Comparison of the time-dependent occupancy $n_d(t)$
           of the resonant-level model, obtain analytically by
           Eq.~(\ref{eqn:nd-T=0}) and numerically by the TD-NRG
           through the evaluation of
           Eq.~(\ref{eqn:time-evolution}). A sudden change of
           $E_d$ from $E_d^0 = 0$ to $E_d^1 = -2\Gamma$ is
           considered without an accompanying change of the
           hybridization strength:
           $\Gamma_0 = \Gamma_1 = \Gamma$. Different values of
           $\Lambda=1.3,1.3^2,1.3^3,1.3^4$ are used. The number
           of NRG iterations (i.e., the Wilson chain lengths)
           have been adjusted so that all curves are calculated
           at approximately the same temperature $T_N$. Explicitly,
           in going from $\Lambda=1.3$ to $\Lambda=1.3^4$,
           $T_N/\Gamma$ equals $0.00171, 0.00176, 0.00183,0.00193$,
           corresponding to $N = 96, 48, 36, 24$ NRG iterations.
           The remaining NRG parameters are as follows:
           $N_s = 1000$, $N_z = 16$, $D/\Gamma = 500$, and
           $\alpha_d = 0$ (i.e., no additional damping).
  }
  \label{fig:res-level-lambda}
\end{figure}

We apply the same technique to improve on the computation
of time-dependent quantities. To partially mimic the
relaxation in an infinite-size system, we calculate the
time evolution of Eq.~(\ref{eqn:time-evolution}) for each
value of $z_i = i/N_z$, $i = 1,\cdots, N_z$, and average
over the different $z_i$'s. Here $N_z$ is the total number
of $z$-values considered. Since the NRG oscillations
are primarily associated with $\sin(2\pi z)$ and
$\cos(2\pi z)$ terms on the interval $0 < z \leq
1$,~\cite{YoshidaWithakerOliveira1990} $N_z$ should
be chosen in multiples of $4$. This leads to optimal
cancellation of oscillations. As shown below (see
Fig.~\ref{fig:res-level-lambda}), usage of the
$z$-trick can greatly improve on the long-time behavior
of the TD-NRG.

Even with the $z$-trick at hand, the evaluation of
$O(t)$ boils down to summation over a finite number of
oscillatory terms of the form $e^{i(E_{r}^m - E_{s}^m) t}$.
One way to simulate a continuous spectrum is to broaden the NRG
levels, as is done in the calculation of equilibrium
spectral functions. This requires, however, extra care
in the nonequilibrium case. As mentioned above, the
limit $O(t \to \infty)$ originates from the
time-independent terms with $E_{r}^m = E_{s}^m$ in
Eq.~(\ref{eqn:time-evolution}). These terms must remain
in tact in order to correctly describe the long-time
behavior. Therefore, one must separate the time-independent
terms from the oscillatory ones. Damping should only
enter the latter terms in Eq.~(\ref{eqn:time-evolution}).

Focusing on the oscillatory terms, we replace each
energy $E_{s}^m$ with a Lorentzian broadening
according to
\begin{equation}
e^{\pm i E_{l}^m t} \to \int \frac{dE}{\pi}
       \frac{\Gamma_m}{(E - E_{l}^m)^2 + \Gamma_m^2}
       e^{\pm i E t} = e^{\pm i E_{l}^m t - \Gamma_m t}.
\label{eqn:Lorentzian}
\end{equation}
Here $\Gamma_m = \alpha_d D_m$ with $\alpha_d$ a
constant of order one is a scale-dependent broadening,
reflecting the characteristic energy resolution of the
$m$th NRG iteration. In this fashion, each of the terms
$E_r^m \neq E_s^m$ in Eq.~(\ref{eqn:time-evolution})
is modified according to
\begin{equation}
e^{i(E_{r}^m - E_{s}^m) t} \to
        e^{i(E_{r}^m - E_{s}^m) t} e^{ -\alpha_d D_m t} ,
\end{equation}
while the terms with $E_r^m = E_s^m$ are left in tact.
Below we present results with and without the additional
damping factor $\alpha_d$.

It should be noted that the Lorentzian broadening of
Eq.~(\ref{eqn:Lorentzian}) differs from the customary
log-normal form used for equilibrium spectral functions.
This choice is motivated by physical considerations, as
it produces a simple exponential decay in time. We also
stress that $\alpha_d$ comes to simulate the continuous
spectrum of the bath. Should ${\cal H}$ possess
eigenstates which are pure eigenstates of the impurity
part of the Hamiltonian ${\cal H}_{\rm imp}$ and do not couple to the
bath (as is the case in certain parameter regimes of electron-transfer
models~\cite{TornowBulla2005}), $\alpha_d$ must be set
to zero. Else, unphysical damping is introduced.

\subsection{Discussion of the TD-NRG approach}
\label{sec:td-nrg-discussion}

For clarity, we summarize the key ingredients as well as the
assumptions made in the TD-NRG approach in this section. We have to
divide the different aspects of conceptual and technical nature.

As mentioned in the previous section, one of the most important
conceptual points to bear in mind is the mimicking of a continuum of
bath degrees of freedom by a selection of discrete states. Therefore,
we expect always some oscillatory contributions in the time
evolution of $O(t)$: the shorter the chain length, the smaller the
amount of bath degrees of freedom the stronger the oscillations. This
property is inherent to any discrete representation of a continuum and
not a shortcoming specific to our method. In fact, the mapping error is
controlled by the NRG discretization parameter $\Lambda$ and the chain
length becomes infinite for a given temperature $T$ for $\Lambda\to
1^+$. In addition, we could show that these oscillations are strongly
suppressed by the $z$-trick described in the previous section also
mimicking a bath continuum.

Another conceptual point concerns our expectation of the time
evolution. If we would include all possible physical interactions, have
all baths  for energy and particle exchange coupled to the system,
wait an infinitely amount of time, we expect that the system
equilibrates to the new thermodynamic equilibrium governed by ${\cal
 H}^f$ independent of the initial conditions. This is required by the basic
assumption of equilibrium thermodynamics. However, this is in general
not the case. Imagine an isolated spin in metallic host in strong
magnetic field, as we will investigate later in the section on the
Kondo model. If we artificially decouple this spin from the
environment and switch of the external magnetic field, the local spin
will remain its  polarized steady state rather than relaxing to the
new thermodynamics state since the local spin remains a conserved
quantity. The local expectation value  can only evolve towards the new
thermodynamics equilibrium, if ${\cal H}^f$ provides an energy,
magnetization or particle exchange mechanism. 

Imagine, $\Delta \H$ is controlled by an external time-dependent field $f(t)$
which couples to a conserved quantity $Q$ of the total system, i.~e.\
$[\hat Q, \H^i] = 0$ and $\Delta \H = f(t) \hat Q$. Then all
eigenstates of $\H^i$ remain eigenstates to $\H^f$ for arbitrary
times. Only the eigen-energies obtain a time-dependent shift $E_m(t) =
E_m^i + f(t) q$, where $q$ is the eigenvalue of $\hat Q$ of state
$\ket{m}$. It is obvious that there will be no time dependence of any
operator whose matrix elements are energy-independent such as spin or
charge.

Another non-trivial difference to the assumption of  thermodynamics
arises from the fact that we describe the time-evolution of a {\em
  closed} system. This is in contrast to the Keldysh approach
\cite{Keldysh65} which introduces - usually in a very subtle -  an
infinitely small relaxation rate much smaller than any energy scale in
the problem to ensure the asymptotic approach of the thermodynamical
state. Sometimes, even explicitly the existence of an additional
thermodynamic bath is made. In our description, however, the total
energy of the system remains a constant and is given by
\begin{eqnarray}
  \label{eq:energy expecation}
  \expect{{\cal H}}(t\ge 0) &=&  {\rm Tr}\{ \hat \rho(t) {\cal H}(t) \}
={\rm Tr}\{ \hat \rho_0 {\cal H}^f \}
\end{eqnarray}
since the time evolution operator $\exp(-i\H^f t)$  commutes with
${\cal H}^f$. In 
contrary, the total energy of the thermodynamical equilibrium is given by
\begin{eqnarray}
 \expect{{\cal H}^f}  & =& {\rm Tr}\{ \hat \rho_f(\beta)  {\cal H}^f \}
\end{eqnarray}
where $\hat \rho_f(\beta) = \exp(-\beta  {\cal H}^f) /{\rm Tr} \{ \exp(-\beta
{\cal H}^f ) \}$. 
A sudden change in the Hamiltonian of closed finite size system will
always result in an effective heating, where the effective temperature
$\beta_{eff}$ can be obtained by solving the implicitly equation
\begin{eqnarray}
{\rm Tr}\{ \hat \rho_0(\beta) {\cal H}^f \} &=&  
{\rm Tr}\{ \hat \rho_f(\beta_{eff})  {\cal H}^f \} 
\,\, .
\end{eqnarray}

If the energy of the  bath states  of  ${\cal H}^{i/f}$, however,
are continuously distributed  and their energies are not changed by
$\Delta \H$,  the temperature will not be changed  in the thermodynamic
limit.

\section{Fermionic benchmark: Resonant level model}
\label{sec:rlm}

\subsection{Definition of the model}

The resonant-level model is perhaps the simplest
example for a quantum-impurity system. A single Fermionic
level $d^{\dagger}$ with energy $E_d$ is coupled by
hybridization to a bath of spinless Fermions:
\begin{eqnarray}
{\cal H} &=& \sum_{k} \epsilon_k c^\dagger_k c_k
          + E_d(t) d^\dagger d
\nonumber \\
&& + V(t) \sum_k
          \left \{
                c^\dagger_k d + d^\dagger c_k
          \right \} \; .
\end{eqnarray}
Here $c^\dagger_k$ creates an electron with momentum $k$
in an s-wave state about the position of the level (taken
to be the origin). To make connection to the TD-NRG
approach, we consider a step-like change in $E_d(t)$ and
$V(t)$: $E_d(t) = E_d^0 \theta(-t) + E_d^1 \theta(t)$ and
$V(t) = V_0 \theta(-t) + V_1 \theta(t)$. Since the model
is bilinear in Fermionic operators, equilibrium properties
(i.e., for $E_d^0 = E_d^1 = E_d$ and $V_0 = V_1 = V$) can
be calculated exactly using, for instance, the local Green
function $G_d(z)$:
\begin{equation}
G(z) = \frac{1}{z - E_d - \Delta(z)} \; ,
\end{equation}
\begin{equation}
\Delta(z) = \sum_k \frac{|V|^2}{z - \e_k} \; .
\end{equation}
In the wide-band limit, $\Gamma = {\rm Im} \{ \Delta(-i0^+)\}$
defines the width of the level due to the coupling to the
Fermionic bath.

\subsection{Analytical Keldysh solution for nonequilibrium}

Out of equilibrium, the occupancy of the level, $n_d(t) =
\langle d^{\dagger}(t) d(t) \rangle$, follows directly
from the knowledge of the equal-time lesser Green function
$G^<_d(t,t') = \langle d^\dagger(t') d(t) \rangle$. Using
the Keldysh technique,~\cite{Keldysh65,LangrethWilkins1972}
we derived an exact analytical expression for $n_d(t)$ in
the wide-band limit, for a step-like change of parameters
at $t = 0$:
\begin{equation}
n_d(t) = \rho_F \int_{-\infty}^{\infty}
              f(\epsilon) |A(\epsilon,t)|^2\, d\epsilon
\label{eqn:nd-general}
\end{equation}
with
\begin{eqnarray}
A(\epsilon,t) &=& \int_{-\infty}^t d\tau \, V(\tau)
                  e^{-i\epsilon\tau - i\int_{\tau}^t d\xi
                       \left[
                            E_d(\xi) + i \Gamma(\xi)
                       \right] }
\nonumber \\
&=& e^{-i\epsilon t}
         \frac{V_1}{\Gamma_1 + i(E_d^1 -\epsilon)}
+ e^{-(i\epsilon +\Gamma_1)t}
\nonumber \\
&&\times
\left[
      \frac{V_0}{\Gamma_0 + i(E_d^0 -\epsilon)}
      - \frac{V_1}{\Gamma_1 + i(E_d^1 -\epsilon)}
\right] \; .
\label{eq:A(e,t)}
\\
\nonumber 
\end{eqnarray}
Here, $\rho_F = \sum_k \delta(\epsilon_k)$ is the density
of states of the Fermionic bath, $f(\epsilon)$ is the
Fermi-Dirac distribution function, $\Gamma(t)$ equals
$\pi \rho_F |V(t)|^2$, and $\Gamma_i = \pi \rho_F |V_i|^2$
($i = 0, 1$).

Besides the wide-band limit, Eqs.~(\ref{eqn:nd-general})
and (\ref{eq:A(e,t)}) require that $|V_0| > 0$, which
comes to ensure that the initially decoupled level for
$t_0 \to -\infty$ has decayed to its equilibrium state
for $-\infty < t < 0$. Since any infinitesimal value of
$V_0$ fulfills this requirement, the limit $V_0\to 0$
can be viewed as contained in Eqs.~(\ref{eqn:nd-general})
and (\ref{eq:A(e,t)}).

For $T \to 0$, Eqs.~(\ref{eqn:nd-general}) and
(\ref{eq:A(e,t)}) can be evaluated in closed analytic
form. Introducing the auxiliary function
\begin{eqnarray}
F(x,y;t) &=& -e^{ty + itx} E_1(ty + itx)
\nonumber \\
&&
          + 2\pi i \theta(-x) \theta(-ty)
                   {\rm sign}(t) e^{ty + itx} ,
\end{eqnarray}
where $E_1(z)$ is the exponential integral function,
we obtain
\begin{widetext}
\begin{eqnarray}
n_d(t > 0,T=0) &=& \left(1+e^{-2\Gamma_1 t}\right) n_d^1
                   + e^{-2\Gamma_1 t} n_d^0
\nonumber \\
&& -2s\sqrt{\Gamma_0\Gamma_1}  \frac{e^{-2\Gamma_1 t} }{\pi}
{\rm Re}\left[ \frac{1}{E_d^0 -E_d^1 - i(\Gamma_0+\Gamma_1)}
               \left[
                      \ln(E_d^0 - i\Gamma_0) -
                      \ln(E_d^1 + i\Gamma_1)
               \right]
        \right]
\nonumber \\
&& +2s\sqrt{\Gamma_0\Gamma_1}  \frac{e^{-\Gamma_1 t} }{\pi}
{\rm Re}\left[
               \frac{e^{-iE_d^1 t}}
                    {E_d^0 - E_d^1 - i(\Gamma_0+\Gamma_1)}
               \left[
                      F(E_d^0,\Gamma_0;t) -
                      F(E_d^1,-\Gamma_1;t)
               \right]
        \right]
\nonumber \\
&& + 2 \frac{e^{-\Gamma_1 t} }{\pi}
{\rm Im} \left[ e^{-iE_d^1 t}
                \left[
                       F(E_d^1,\Gamma_1;t) -
                       F(E_d^1,-\Gamma_1;t)
                \right]
         \right] \; .
\label{eqn:nd-T=0}
\end{eqnarray}  
\end{widetext}
Here $n_d^i = \frac{1}{2} - \frac{1}{\pi}
\arctan(E_d^i/\Gamma_i)$ with $i = 0, 1$ is
the equilibrium occupancy of the level for
the corresponding model parameters, and
$s = {\rm sign}(V_1 V_0)$.

We note that the exact result does not exhibit a simple
exponential decay to the new equilibrium occupancy $n_d^1$.
It is actually governed by two relaxation rates, $\Gamma_1$
and $2\Gamma_1$. Moreover, the last and second-to-last terms
in Eq.~(\ref{eqn:nd-T=0}) contain an oscillatory factor
$\exp(-iE_d^1 t)$, responsible for Rabbi-type oscillations
that visible for $|E_d^1|/\Gamma_1> 1$ at short time scales.
In a previous publication,~\cite{AndersSchiller2005} we
have used this rather elaborate formula and its extension
to finite temperatures to benchmark the TD-NRG. Here we
complete the discussion by focusing on the role of the
NRG parameters $\Lambda$ and $N_z$.

Figure~\ref{fig:res-level-lambda} presents a comparison
between the exact analytic result of Eq.~(\ref{eqn:nd-T=0})
and the time-dependent occupancy $n_d(t)$ obtained by the
TD-NRG for different values of $\Lambda$ and $N_z$. The
temperature is roughly kept fixed in all curves by
adjusting the length of the NRG chain. While the curve
corresponding to the smallest value of $\Lambda$ and,
thus, to the longest chain used ($N=96$) shows excellent
agreement with the exact analytical result on all time
scales, the curves for the larger values of $\Lambda$
show good agreement only at shorter times, $t\ \Gamma < 2$.
Deviations from the exact result develop at longer
times, which are characterized by oscillations about
a value of $n_d$ that is reduced as compared to the new
thermodynamic average. We emphasize that none of the
TD-NRG curves in Fig.~\ref{fig:res-level-lambda}
involved an extra damping factor $\alpha_d > 0$.

The above behavior is not surprising since, as discussed
in Sec.~\ref{sec:td-nrg-discussion}, the shorter the NRG
chain, the worse it represents a continuous bath. A different
way of viewing the matter is through the conservation of
energy and particles in the system for $t > 0$. In a
macroscopic bath, the system relaxes to the new
equilibrium state by redistributing the excess particles
among the different lattice sites. Since the excess number
of particles is of order one, each lattice site (the level
included) acquires an infinitesimal shift to its occupancy,
which vanishes in the thermodynamic limit. For a finite
system, the shift in occupancy remains finite. We therefore
conclude that the oscillations depicted in
Fig.~\ref{fig:res-level-lambda} are real in a finite-size
system. However, they are unphysical for a continuous
bath, as demonstrated by the exact analytical solution. We
pinpoint this ``inaccuracy''  to be of conceptual nature,
controllable by the limit $\Lambda \to 1^+$.

The magnitude of the finite-size oscillations is greatly
reduced, though, by resorting to Oliveira's z-trick. To
illustrate this point, we have added one extra curve for
$\Lambda=1.3^4=2.8561$ without any $z$-averaging. Enhanced
by the short chain length, the finite-size features are
notably more noisy and have a larger amplitude. The effect
of averaging over $z$ is to greatly smoothing and reduce these
finite-size structures. Hence averaging over $z$ is vital
for describing the long-time behavior, particularly if no
extra damping is introduced.

The short time behavior, however, is  very accurately reproduced
defying  prejudice against the NRG which allegedly is insufficient to
describe the high energy physics correctly needed for the short time
behavior. Even if the statement would be true, the NRG represents the
high energy physics incorrectly  for calculating 
thermodynamics properties, which is not --  see Wilson
\cite{Wilson75} -- it would have no influence on the time evolution
calculated with the TD-NRG for a very simple reason. The high energy
states are generated at the first iterations. For energy scales much
larger than the relevant scales of the quantum impurity, the NRG flow
of ${\cal H}^i$ and ${\cal H}^f$ are more or less identical which
leads to a almost diagonal overlap matrix $S_{r,s}(m)$ for these
iterations. Since the reduced density matrix $\rho^{\rm red,0}_{k,k'}(m)$
is only non-zero for retained states $k,k'$ there will be no
contribution to the time evolution when evaluating (\ref{eqn:time-evolution}).
Only when the flow between the two Hamiltonians starts to deviate,
contributions to the expectation value $O(t)$ are generated. This may
be seen clearly in the inset of figure 2a of our previous paper
\cite{AndersSchiller2005} which shows the iteration dependent
evolution of $n_d(t\to 0^+)$.

\section{Bosonic benchmark: decoherence in the spin-boson model} 
\label{sec:spin-boson}

The spin-boson model \cite{Leggett1987,Weiss1999}
\begin{eqnarray}
  H&=& -\frac{\Delta}{2}\sigma_x + \frac{\e}{2}\sigma_z
+ \sum_i \w_i a^\dagger_i a_i 
\nonumber \\
&& + \frac{\sigma_z}{2}\sum_i
\lambda_i(a^\dagger_i + a_i)
\end{eqnarray}
is perceived as the simplest model for the understanding of dissipation
and decoherence in quantum systems. A two-level system, represented by
a spin, is coupled to the displacement operator of a continuum of
bosonic degrees of freedom with the coupling constant $\lambda_i$ to
each bosonic mode. This continuum causes decoherence, dissipation
and possible spin decay. The effect of 
this dissipative environment is fully captured by the coupling
function $J(\w)$  
\begin{eqnarray}
  J(\w) &=& \pi \sum_i \lambda^2_i\delta(\w-\w_i) \; .
\end{eqnarray}
Since the low-energy part of the spectrum dominates the response of
the spin, the  high-energy details of $J(\w)$ can be neglected: this
function is usually approximated by a power-law behavior \cite{Leggett1987}: $J(\w) \propto
\w^s$, $0<\w\le w_c$. In the literature, one distinguishes between the
``ohmic'' case, $s=1$, corresponding to a classical model with a
dissipative term proportional to the velocity, the ``sub-ohmic'' case
$0\le s<1$ and the ``super-ohmic'' case $s>1$.
In the ohmic regime, the spin-boson model can be mapped
onto the anisotropic Kondo model in the wide-band
limit,~\cite{Leggett1987} with $\Delta/\w_c=\rho_F J_\perp$
and $\alpha=(1-\rho_F J_z)^2$.

This model plays also an important role for understanding charge
transfer reactions in chemistry \cite{XuSchulten1994} where the spin
states would represent the initial and final state of the transfer reaction
and the bosonic continuum the density fluctuations of solvents or
molecular vibrations. It has been argued
\cite{XuSchulten1994,Schulten2000} that the  model serves a quantum
mechanical foundation of the classical Marcus theory \cite{Marcus1993}
for electron transfer reaction.

In addition, the model has been  discussed in the context of decoherence of
qubits in quantum computing
\cite{Unruh1995,PalmaSuominenEkert1996}. Setting $\Delta=0$,
$\sigma_z$ is a conserved quantity, and, therefore, the bosonic
environment can be traced out analytically. If we prepare an initial
density matrix containing locally only the pure 
state $\ket{s} = (\ket{0} + \ket{1})/\sqrt{2}$, it was shown
\cite{PalmaSuominenEkert1996} that off-diagonal part of the reduced
density matrix  
\begin{eqnarray}
\label{eqn:rho-10}
  \rho_{10}(t) &= & \bra{1}{\rm Tr_{Boson}} \{ \rho(t) \} \ket{0}
\end{eqnarray}
can be written as $\rho_{10}(t) = e^{-\Gamma(t)} \rho_{10}(0)$. Here,
$\Gamma(t)$ is given by the exact expression 
\begin{eqnarray}
  \Gamma(t) &=& \frac{1}{\pi} \int_{0}^\infty  \, d\w \, J(\w)
  \coth\left(\frac{\w}{2T}\right) \frac{1-\cos(\w t)}{\w^2}
\end{eqnarray}
for the temperature $T$. \cite{Unruh1995,PalmaSuominenEkert1996}

\begin{figure}[tbp]
  \centering
  \includegraphics[width=85mm]{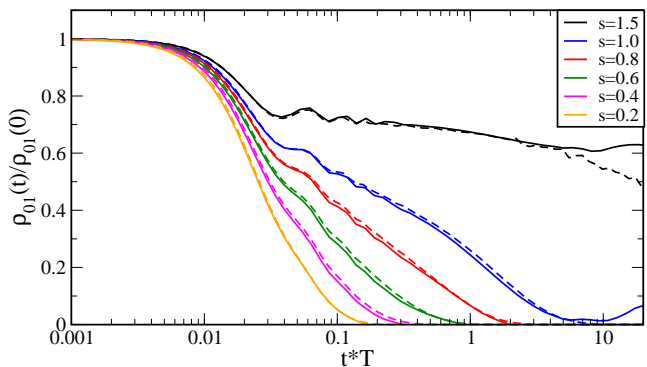} 

  \caption{Decoherence of the off-diagonal density matrix element
    $\rho_{01}$ of the local reduced density matrix in the spin-boson
    model  as function of time, by measuring the expectation value
    $s_x(t) =\rho_{01}$. The solid lines show the TD-NRG 
    results, the dashed line with the same color (online)
    the exact analytical result given by (\ref{eqn:rho-10}). 
    NRG parameters: $N_s=150,N_{iter}=14,
    N_b=8$ (number of bosons added a each iteration
    \cite{BullaBoson2003},) $N_z=16, \Lambda=\sqrt{2},
    T=0.0078,\alpha=0.1,\w_c=1, \Delta^1=\epsilon^1=0,\alpha_d=0.1$. } 
  \label{fig:spin-boson-decoherence}
\end{figure}

Recently, Bulla and collaborators  introduced an extension of the NRG
to bosonic baths.\cite{BullaBoson2003,BullaVoita2005} These authors
investigated the different quantum critical points of the model as a
function of the bath exponent $s$ and the coupling strength $\alpha$. We
combined their bosonic NRG method and our TD-NRG approach to
show explicitly that our method is very accurate for short as well as  long
times up to $t*T\approx O(1)$  for bosonic baths as well.

In order to start with the same boundary conditions as the exact
result was derived for, we set $\Delta^0=100$ in ${\cal H}^i$. This
guarantees that equilibrium density operator consists of the
disentangled operator product of 
$\ket{s}\bra{s}$ and the equilibrium density matrix of the bosonic
bath for $t<0$. For $t>0$, $\Delta$ is set to zero and an entangled
density matrix is evolving with time. The off-diagonal matrix element
$\rho_{10}$ corresponds to the expectation value of $s_x(t)$
obtained by the bosonic TD-NRG. The coupling function $J(\w)$  
\begin{eqnarray}
  J(\w) &=& 2\pi \alpha \w_c^{1-s} \w^s \;\; {\rm for} \; 0<\w \le\w_c
\end{eqnarray}
enters the analytical as well as the numerical calculation. We have
obtained  $\rho_{01}(t)/\rho_{01}(0)$ for fixed 
$\alpha=0.1$  and different exponents $s$. The comparison between the
TD-NRG and the {\em exact analytical result} given by Eq.\
(\ref{eqn:rho-10}) is depicted in Fig.\
\ref{fig:spin-boson-decoherence} for the super-ohmic, the ohmic and the
sub-ohmic case. We observe excellent agreement between the exact
analytical result and the predictions of the TD-NRG.
\footnote{A detailed analysis of the dynamical properties of spin-boson
model will be published elsewhere.} The small oscillations in the
quantum regime  observed in all  curves in the time range $0.01<t * T <1$
are no artifacts of the method but real features. By comparison with
the exact analytical result, they have been traced to the usage of the
hard frequency cutoff in $J(\w)$ at $\w=\w_c$. Using a soft
cutoff function $J(\w)\propto \w^s\exp(-\w/\w_c)$  would produces
smoother and slightly shifted curves.\cite{PalmaSuominenEkert1996} 

We noted that due to the  larger numbers of degrees of freedom added
though  each chain link, the time evolution for a bosonic bath shows a higher
accuracy and less dependence on $\Lambda$ than for fermionic baths as
long as one does not leave the range of validity of the bosonic chain
NRG. \cite{BullaVoita2005}

\section{The Kondo model: analytical considerations}
\label{sec:kondo-model-analytics}

\subsection{Definition of the model}

The Kondo model comprises of a local spin $\vec{S}_{imp}$ interacting
with a spin degenerate conduction band via a local anisotropic
Heisenberg term  
\begin{eqnarray}
\label{eqn:kondo-ww}
  H_{K} &=& J_z \sum_{\sigma} \sigma
  c^\dagger_{R=0,\sigma}
c_{R=0,\sigma}  S^z_{imp} 
\\
&& \nonumber
+ J_\perp
\left(
  c^\dagger_{R=0,\uparrow} c_{R=0,\downarrow}  S^-_{imp} 
+
  c^\dagger_{R=0,\downarrow} c_{R=0,\uparrow}  S^+_{imp} 
\right)
\end{eqnarray}
where  $c^\dagger_{R=0,\sigma}$ creates a conduction electron at the
position of the spin $R=0$. For $J_z=J_\perp$, we obtain the usual
$SU(2)$ symmetric interaction. We also allow for a time-dependency of
the coupling constants $J_z$ and $J_\perp$. The energy of the local levels are
splitted in an external magnetic field $\vec{H}(t)$ by the Zeeman energy
\begin{eqnarray}
  H_{loc} &=& - \gamma_s \vec{H}(t) \vec{S}_{imp}
\label{eqn:zeemann}
\end{eqnarray}
whose transversal parts can also be interpreted as hopping terms of
the two-level system modelled by the local spin 1/2.
The local conduction electron creation operators are
expanded in s-wave eigenmodes  of the non-interacting conduction electron band
\begin{eqnarray}
   c^\dagger_{R=0,\sigma} &=& \int_{-\infty}^\infty d \e \, \sqrt{\rho(\e)}
   c^\dagger_{\e\sigma}  
\end{eqnarray}
where $\rho(\e)$ denotes the conduction electron density of states.
\footnote{Even though the
formulation is quite general, we will use a constant density of states
$\rho(\e)=1/(2D)$ and a symmetric band $D_{up}=-D_{low} =D$ for all
numerical calculations throughout the paper.} Note that the anti-commutator
of the local conduction electron operator is given by $\{
c^\dagger_{R=0,\sigma}, c_{R=0,\sigma'} \} = \delta_{\sigma\sigma'}$.
The conduction electron Hamiltonian is given by
\begin{eqnarray}
  H_{c} &=& \sum_\sigma \int_{D_{low}}^{D_{up}}
 d\e \, \e \, c^\dagger_{\e\sigma}c_{\e\sigma}
\end{eqnarray}
where $c^\dagger_{\e\sigma} (c_{\e\sigma})$ creates (annihilates) 
electron in an s-wave state with spin $\sigma$, energy $\e$.  All
other angular momentum do not interact with the local 
spin.\cite{Wilson75} The integration runs from the lower  to the upper
band edge, $D_{low}$ to $D_{up}$. Throughout the paper, we assume a
symmetric band with a relativistic dispersion, i.\ e.\ $D_{low}
=-D_{up}=-D$ and $\rho_F= \rho(\e)=const=1/(2D)$.  
Then, the total Hamiltonian under consideration  is given by
\begin{eqnarray}
{\cal   H}(t) &=& H_{c} + H_{K}(t) +  H_{loc}(t)
\; .
\end{eqnarray}

Why do  we neglect the spin polarization of the
conduction electrons? As we discussed in section
\ref{sec:td-nrg-discussion}, an external field coupling to total
$z$-component of the spin yields a time independent local spin
expectation.  Since the external magnetic field, however, couples to
the total magnetization 
\begin{eqnarray}
  \vec{M}_{tot} &=& \gamma_{s} \vec{s}_{s} + \gamma_{b}
  \vec{S}_{b}^{tot} 
\nonumber \\
&=& \left(\gamma_s -\gamma_b\right)\vec{s}_{s}  + \gamma_b\vec{S}^{tot}
\end{eqnarray}
where $\vec{s}_z$ is the local spin operator and $\vec{S}_b^{tot}$
the total spin of the conduction electrons, the relevant Zeeman energy
for the real time dynamics is given by 
$H_{zee}= - \left(\gamma_s  -\gamma_b\right)\vec{s}_{s}\vec{H}$. 
In this case, the local spin can 
relax  if the gyromagnetic ratios of local spin $\gamma_s$ and
conduction electrons $\gamma_b$ differ. Then, $\gamma_s$  would have
to be replace by $\gamma'_s=(\gamma_s -\gamma_b)$ in Eq.\
(\ref{eqn:zeemann}).

In real materials, however, the spin-orbit scattering of the
conduction electrons on distributed impurities in the metallic host
furbishes another relaxation mechanism, as pointed out by Langreth
and Wilkins \cite{LangrethWilkins1972} more than 30 years ago. We
might argue in the following way. If this 
relaxation processes for the conduction electron magnetization is much
faster than the relaxation process of the impurity spin due to
coupling $J$ to the conduction electrons, we
can consider the conduction electron band as demagnetized on the
time scale of the impurity relaxation  process. In this case, 
neglecting the spin-polarization of the conduction electrons are justified.
If for some reason, the spin-lattice relaxation processes is very slow
compared to the impurity spin, the impurity spin-relaxation process is
dominated by the spin-lattice relaxation time in which case we have to
substituted $\gamma'_s=(\gamma_s -\gamma_b)$ or rescale the absolute
values of the physical magnetic field.

It turns out, that the long-time behavior of impurity spin relaxation
is dominated by the Kondo time $t_K$ in the Kondo regime, given by the
reciprocal characteristic thermodynamic energy scale $t_K=1/T_K$, and
by the reciprocal temperature $1/T$ for the ferromagnetic regime of
the model. As long as the spin-lattice relaxation time
is finite, for sufficiently low temperature $T$ and Kondo temperature
$T_K$, neglect of the conduction electron spin-polarization is
justified. However, the short time behavior of the spin relaxation 
has admittedly purely academical value if $t_{sp}> 1/J$.

We will distinguish between two different starting positions. Either
we leave the coupling constant unaltered and just switch of the
magnetic field at $t=0$, which is the physically more relevant
condition. Alternatively, we start  from an initially decoupled
system, the local moment fixed point of the Hamiltonian, and switch on
the Kondo couplings $J_z$ and $J_\perp$ as well as changing the
applied magnetic field at time $t=0$. It was 
previously pointed out \cite{NordlanderEtAll1999,AndersSchiller2005}
that this setup is closer to the thermodynamics since the time $t$ might play a
similar role as $\beta$.

\subsection{Analytical results}
\label{sec:analytic}

Wilson invented the numerical renormalization group\cite{Wilson75}
to solve the thermodynamics of this model very accurately for all
parameter regimes. The 
model has two fixed points for the antiferromagnetic regime, the
unstable local moment fixed point $(J_Z=0,J_\perp=0)$ corresponding to
$\beta= 1/T\to 0$ and the stable strong coupling fixed point $(J_z\to
\infty,J_\perp\to \infty)$ for $\beta\to\infty$ caused by the infrared
divergence in the perturbation theory in the absence of an external
magnetic field. For the 
ferromagnetic regime, i.e.\ $J_z<0$ and $|J_z|\le |J_\perp$, the stable
fixed point is given by $(J_z(\infty) 
<\infty,J_\perp  =0)$. Due to the ferromagnetic polarization of the
conduction band in the vicinity of the impurity, the spin-flip
scattering is more and more suppressed for decreasing temperature:
the model remains perturbative in this regime. 

The analytical ``poor man's scaling'' second-order renormalization group
treatment\cite{Anderson70} prior to the NRG predicts
\begin{eqnarray}
  J^2_z - J_\perp^2 &=& {\rm const.} 
\end{eqnarray}
The renormalization of isotropic coupling $J(D')$ as of the
running cutoff $D'$
\begin{eqnarray}
  J(D') &=& \frac{J(D)}{1 - J\ln(D/D')}
\label{eqn:running-J}
\end{eqnarray}
provides some  analytical inside for the fixed point structure
even though the treatment breaks down for $J\ln(D/D')\to 1$.

For the isotropic ferromagnetic Kondo model, however, $J(T)$ remains
finite and vanishes for $T\to 0$. Since the model approaches the local
moment fixed point in this case, a spin-polarized local state remains
spin-polarized after switching of the magnetic field at $T=0$ without
any additional relaxation mechanism present. At finite temperature,
however, a thermal energy of $T$  provided fluctuations in the bath in
addition to the very small but finite transversal spin flip term
$J_\perp(T)=J(T)$. Therefore, we expect that the FM Kondo model
has a characteristic time scale $t_{FM} = |2 \rho_F J(T) T|^{-1}$ which
is actually found in the  calculations ($\hbar=1$).
Through the paper we measure the temperature in energy units ($k_B=1$.)

\subsection{Short time behavior and relaxation towards the Kondo 
  strong coupling fixed point}
\label{sec:analytical-short-time}

If we consider a decoupled impurity for times $t<0$, we
can calculate analytically the short time response  using a
perturbation expansion. For $J_z^0=J_z(t<0) =0$ and 
$J_\perp^0=J_\perp(t<0)=0$, the equilibrium density operator is given
by $\exp(-\beta H_0^0)/Z$ where $H_0^0 = H_c +H_{loc}$. Switching on
the coupling $J_z^1=J_z(t>0)$,$J_\perp^1=J_\perp(t>0)$ at $t=0$ leads
to the time evolution of spin operator 
\begin{eqnarray}
  \label{eq:time-evolution-sz}
  S_z(t) &=& e^{iH t}S_z e^{-iHt}
\end{eqnarray}
where $H=H_c + H_K$, if the magnetic field is switched off for
$t>0$. In the interaction picture, the spin operator 
\begin{eqnarray}
    S_z^I(t) &=& e^{-iH_c t}S_z(t) e^{iH_c t}
\end{eqnarray}
obeys the equation of motion
\begin{eqnarray}
   \frac{\partial S_z^I(t)}{\partial t} &=& 
i[ H_K^I(t),S_z^I(t)]
\end{eqnarray}
which is integrated to
\begin{widetext}
\begin{eqnarray}
  \label{eq:sz-integration}
    S_z^I(t) &=& S_z + i \int_0^t d\tau_1 [H_K^I(\tau_1),S_z(0)]
+ i^2\int_0^t d\tau_1\int_0^{\tau_1} d\tau_2
 [H_K^I(\tau_1), [H_K^I(\tau_2), S_z(0)] ]
\\
&&
\nonumber
+ i^3\int_0^t d\tau_1\int_0^{\tau_1} d\tau_2\int_0^{\tau_2} d\tau_3
 [H_K^I(\tau_1), [H_K^I(\tau_2), [H_K^I(\tau_3), S_z(\tau_3)] ] ]
\end{eqnarray}  
\end{widetext}
where $H_K^I(t) = e^{-iH_c t} H_K e^{iH_c t}$. Neglecting the last
term yields a perturbative result correct up to order $O(J^2)$. Since
$S_z$ commutes with the $S_z$ term in $H_K$, its short time behavior
is solemnly governed by $J_\perp$. The renormalization of $J_\perp$ by
$J_z$ sets in at third order,  well known for the Kondo
problem. \cite{Hewson93} Alternatively, the time evolution of
$\expect{\hat S_z(t)}$ could  have also been  calculated using Keldysh
techniques.\cite{LangrethWilkins1972} All expectation values have to
be evaluated with equilibrium density operator $\hat \rho_0$.
Since we are only interest in an analytical results for short times,
the Keldysh approach does not have any advantages: it also fails to
capture the Kondo physics at  long time scales.\cite{LangrethWilkins1972}

There will be no contribution linear in $O(J_z,J_\perp)$ since
$\expect{S^+}=\expect{S^-}=0$. 
Evaluation of the two commutators needed for the second order contribution
$\Delta S_z^{(2)}$ in
(\ref{eq:sz-integration}) yields 
\begin{eqnarray}
  \label{eq:second-order}
\Delta S_z^{(2)}  &=& \expect{S_z}( 2i J_\perp^1)^2
\int_{-\infty}^\infty d\e\int_{-\infty}^\infty d\e'
\rho(\e)\rho(\e')
 \nonumber \\
&& \times f(\e')(1-f(\e))
 \nonumber \\
&& \times
\Re e\left[
\frac{\left(1-e^{-i(\e-\e')t}\right)}{(\e-\e')^2} -
    \frac{it}{\e-\e'}
\right]
\punkt
\end{eqnarray}

For $T\to 0$ and a constant density of states $\rho_F=1/(2D)$,
we can solve the integrals analytically and obtain
\begin{eqnarray}
\label{eqn:sz-analytic-fit}
  \expect{S_z(t)} -\expect{S_z}&\approx&   \Delta S_z^{(2)}
 \\
&=& -\expect{S_z}(2 \rho_F J_\perp^1)^2
\left[G(2Dt) - 2 G(Dt)\right]
\nonumber
\end{eqnarray}
where we have defined the function $G(x)$ by its series expansion
\begin{eqnarray}
  G(x) &=& \sum_{l=1}^\infty \frac{ (-1)^{l+1}}{(2l)! 2l(2l-1)} x^{2l}
\punkt
\end{eqnarray}
The result is exact up to $O(J^2_\perp)$: the short time evolution is
independent of $J_z$, consistent with the physical picture that a
change of the spin state can only by mediated by the spin-flip term of
Kondo interaction. $J_z$ enters at first in the third order of the
perturbation expansion, corrections  which are neglected here.

\section{The Kondo model: numerical results of the TD-NRG}
\label{sec:kondo-model-td-nrg}

\subsection{Antiferromagnetic regime } 
\label{sec:kondo-model-td-nrg-anti}

\subsubsection{Short and long time behavior for $T\to 0$}
\label{sec:short-and-long-time-AF}

\begin{figure}[htbp]
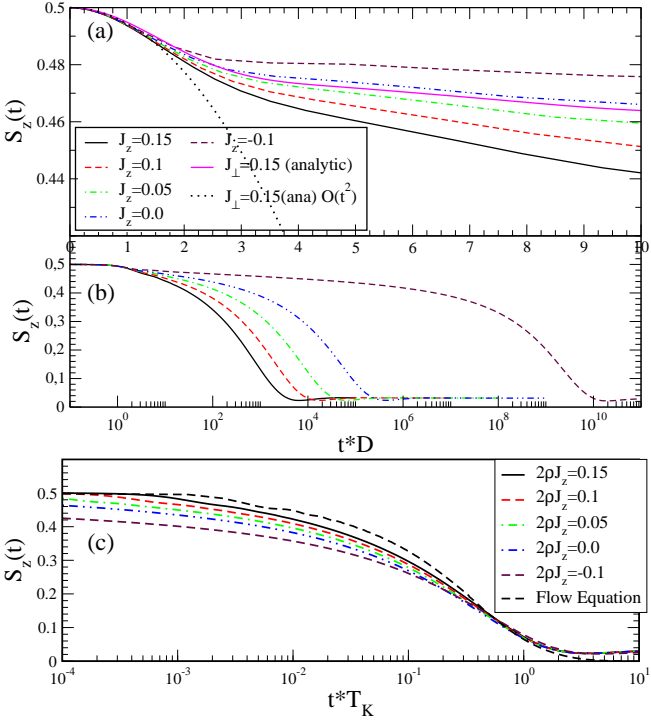

  \centering
  \includegraphics[width=86mm]{anisotrop-Kondo-Jz}

  \includegraphics[width=86mm]{anisotrop-Kondo-Jz-vs-tk}

  \caption{Spin-expectation value for the anisotropic Kondo
    model in the antiferromagnetic regime  vs
    temperature for different values of $J_z/D = -0.1,0.
    0.05,0.1,0.12,0.15$ and fixed $J_\perp^1/D=0.15$.
    The upper panel (a) shows the short time behavior and the analytic
    fit using Eq.\ (\ref{eqn:sz-analytic-fit}) (magenta curve) and
    its the second order contribution in  $t$. The panel (b) displays
    the long time behavior of   the same data  on a logarithmic time
    scale while in (c) the data of (b) is plotted vs $t/t_K$. We
    added the universal curve from Ref.\
    \onlinecite{LobaskinKehrein2005}  where we rescaled their
    definition of $T_K$ by the factor $1/1.67$ to match ours. 
    NRG parameters: $\Lambda=1.5$; $N_s=1000$, $\alpha=0.1$, $N_z=16,
    J^0_z=J_\perp^0 = 0, H_z^0/D=0.1, N_{iter}=130$,$ T=3.7*10^{-12}$.
  }
  \label{fig:af-anisotrop-kondo}
\end{figure}

In order to make connection between the TD-NRG and  the analytical
calculation for the 
short time behavior presented in previous section, we 
start with a decoupled impurity in a finite magnetic field along the
z-axis, i.e.\ $J^0_z=J_\perp^0=0, H_z/D=0.1$.
At $t=0$, we switch on the anisotropic Kondo interaction
and simultaneously switch off the magnetic field. At infinitely
long times, we expect that the system will relax into the strong
coupling Kondo fixed point for $T\to 0$ and $t\to \infty$,
independently of the initial conditions. 

In Fig.\ \ref{fig:af-anisotrop-kondo}, the time evolution of $S_z(t)$
is shown for a fixed spin flip rate $J_\perp/D=0.15$ and different
values of $J_z/D$.  $S_z(t)$  is independent of $J_z$  as expected for
times $t* D\le 1$, clearly visible in the upper panel (a). Only
spin-flip processes can alter the spin polarization. At short times,
the Kondo correlations are absent. Only  the  bare coupling
constants enter the analytical result  calculated for the comparison
with the presented TD-NRG curves. The magenta (online) curve in (a)
displays the second  contribution in $J_\perp$ given by  Eq.\
(\ref{eqn:sz-analytic-fit}) which contains arbitrary high orders in
$t$ through the series expansion of $G(x)$.Since the analytical
formula is  independent of $J_z$, it must correspond to the TD-NRG
curve for $J_z=0$. We find an excellent agreement between the
analytical and  the TD-NRG result confirming the accuracy of our
method at short and intermediate time scales. These findings show that
the ultra-short time response is purely perturbative, since the {\em
  strong correlations} develop only at low energies and, therefore,
influence only the {\em  long time behavior}. The $J_z >0$ curves lie
below, the curve with a ferromagnetic $J_z/D = -0.1$ above the
analytical results. Starting with order $O(J^3)$, $J_z$ terms contribute
to $S_z(t)$ by  renormalizing $J_\perp$. The positive
$J_z$ leads to an increase of $J_\perp$, while a ferromagnetic $J_z$
yields a reduction of $J_\perp$ consistent with our findings in 
Fig.\ \ref{fig:af-anisotrop-kondo}a) of an increasing relaxation time
scale with decreasing $J_z$. The long-time behavior of $S_z(t)$
is displayed in Fig.\ \ref{fig:af-anisotrop-kondo}b). Note that the
TD-NRG accurately predicts the relevant time scales of the spin
relaxation for arbitrary long time scales, here over 10 orders of
magnitude in units of the reciprocal band width $1/D$ as long as 
$t *D \le 1$.

\begin{table}[tbp]
  \centering
  \begin{tabular}{c|c|c}
    $J_z/D$ & $T_K$ & $t_K=1/T_K$ \\
    \hline
    0.15 & 0.000537 &     1862.3 \\
    0.10 & 0.000203941 &    4903.4 \\
    0.05 & $5.49093* 10^{-5}$ &    18211 \\
    0.0 & $8.00952 * 10^{-6}$ &    124850 \\
    -0.1 &  $2.31866*10^{-10}$ &    $4.3128* 10^{9} $\\
    \hline
  \end{tabular}
  \caption{Kondo temperature as function of $J_z$ for fixed
    $J_\perp=0.15$ used
    for rescaling the time axis in Fig.\ \ref{fig:af-anisotrop-kondo}c).
    Note that the Kondo temperature $T_K$, obtained by the thermodynamic
    condition\cite{Wilson75} $\Delta     \expect{S_z^2}(T_K) = 0.07$, 
    has a $\Lambda$  dependence. NRG parameter $\Lambda=1.5, N_s=1000$. } 
  \label{tab:tk-Jz}
\end{table}

In order to identify the long 
time scale, we rescale the data with the thermodynamic Kondo
temperature $T_K$ which is obtained from the temperature dependency of
the effective local moment \cite{Wilson75} $\Delta
\expect{S_z^2}(T_K)=0.07$. Excellent scaling is found for the long
time tails. Instead of using the thermodynamic Kondo temperature, we
could have defined  an effective time scale $t_{low}$ by which
$\expect{S_z}(t_{low})=0.1$ is reduced to 20\% of the starting value
$S_z(0)=0.5$. It turns out that $t_{low} * T_K=const\approx 0.76$
independent for $J_z$ within the numerical accuracy. Therefore, the
two time scales are equivalent, and the Kondo time $t_K=1/T_K$ governs
the long time behavior of $S_z(t)$.

From this discussion, it is apparent that there will be no universal
master curve for $S_z(t)$ valid for all times and  solemnly parametrized by
a universal time scale $t_{low}$. Only at very long times $t\gg
t_{low}$, the curves tent to approach a universal curve
$S(t/t_{low})$. The panel (c) of Fig.\ \ref{fig:af-anisotrop-kondo}
presents how  the value of $J_z$ determines this approach to  this
master curve. We also added the universal curve for 
the weak coupling isotropic Kondo model, i.e.\ $\rho_F J\ll 1$. Lobaskin
and Kehrein have calculated this curve\cite{LobaskinKehrein2005} using 
Wegner's flow equation \cite{Wegner1994} and expanding the solution at
the exactly solvable Toulouse point. The quantitative agreement
between their result and the TD-NRG for isotropic Kondo model is very
good. However, our calculations also show that the transients towards
the ultra long time behavior $t/t_K \to \infty$ at time scale
$t\approx t_K$ are still dependent of the anisotropy ratio
$a=J_z/J_\perp$. Deviations from $a=1$ lead to deviations  from an
master curve for an isotropic Kondo model. Only at $t\to\infty$, we
expect the merging of the curves.

\begin{figure}[tbp]
  \centering
  \includegraphics[width=86mm]{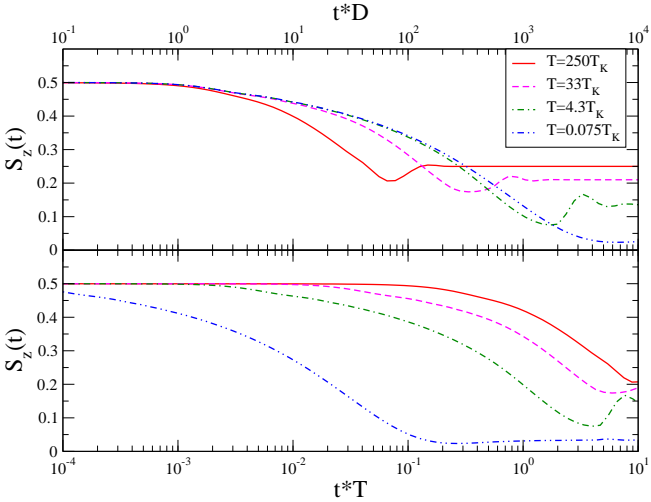}

  \caption{Spin-expectation value $S_z(t)$ for the isotropic Kondo
    spin  with fixed $J/D=0.15$ for different
    temperatures $T=250,33,4.3,0.075 T_K$
    and an initial magnetic field of $H^0_z/D=1$ and $J^0_z=J^0_\perp=0$
    where $T_K/D=0.00054$. The lower panel (b) displays the same data
    as in (b) but plotted as function  of $t*T$.
    NRG parameters: as in Fig.\ref{fig:af-anisotrop-kondo}.
  }
  \label{fig:finite-T-AFM-kondo}
\end{figure}

\subsubsection{Finite temperature relaxation}

Starting from a decoupled, fully spin-polarized impurity, the
temperature dependence of the spin relaxation is plotted in Fig.\
\ref{fig:finite-T-AFM-kondo} for the isotropic, antiferromagnetic
Kondo spin. The characteristic spin-relaxation 
time $t_{low}$ saturates for temperatures $T<T_K$ and is of the order
of $1/T_K$ at very low temperatures, similar to the findings in the
single impurity Anderson model. \cite{AndersSchiller2005}
The upraise of $S_z(t)$ for very long times at the temperatures $T>T_K$ is an
artifact of the insufficient  time resolution at high temperatures. To
illustrate this point, the same data is plotted as function of $t*T$
in the lower panel (b).

\subsubsection{Magnetic field dependence of the spin relaxation}
\label{sec:magnetic-field-dependence}

\begin{figure}[htbp]
  \centering
  \includegraphics[width=86mm]{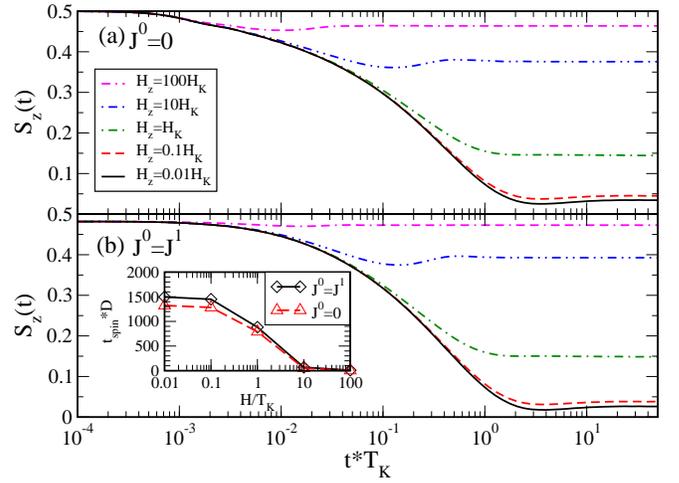}

  \caption{Spin relaxation for the isotropic Kondo
    spin with fixed $J/D=0.15$ for different
    strength of the final magnetic field $H^1_z=0.01,0.1,1.0,10,100 T_K$
    for $T=1.2*10^{-8}$ starting from $H^0_z/D=0.1$. The upper panel
    (a) shows  the case of $J^0=0$, (b) the case $J^0=J^1$. The inset
    in (b) display an estimate for a spin-relaxation scale $t_{spin}$
    as defined in the text.
    NRG parameters: as in Fig.\ \ref{fig:af-anisotrop-kondo}.
  }
  \label{fig:finite-H-kondo}
\end{figure}

The time evolution of a polarized decoupled Kondo spin with
Hamiltonian with a remaining finite local magnetic
field $H^1$ is plotted in Fig.~\ref{fig:finite-H-kondo}. For $t\le 0$ the
local spin is fully polarized by a strong magnetic field. The
figure shows the evolution of $S_z(t)$ for five different magnetic
field strength $H^1 = 0.01,0.1,1,10,100T_K$, where the Kondo scale is
given by $T_K=0.000534 D$ for $J/D=0.15$. For the upper panel (a), the
local spin was decoupled from the bath for times $t<0$, while the
curve in the lower panel (b) were calculated for a Kondo spin coupled
to the bath at all times. The difference in the time evolution between
the two scenarios are very small in contrary to our previous
study of the single impurity Anderson model
\cite{AndersSchiller2005}. The strong  magnetic field destroys the
Kondo correlation  and the regimes are described by a spin polarized
local moment fixed point in both cases: the Kondo spin-flip scattering
is an irrelevant  operator due to the external field. In the Anderson
model, however, the fixed points of both scenarios are completely
different: in case of an absent hybridization, we have a free orbital
or local moment fixed point while for finite coupling we started from  a spin
polarized mixed valent fixed point.

Fig.\ \ref{fig:finite-H-kondo} clearly shows a decrease of the
characteristic time scale with increasing final magnetic field. In
order to quantify this behavior, we asked how long does it take until
the spin-expectation value reaches some fraction of its  asymptotic
value $S_z(\infty)$. For this purpose, we introduce the reduced spin
expectation value function  
\begin{eqnarray}
f(t) &=  & \frac{S_z(t) -S_z(\infty)}{S_z(0) -S_z(\infty)}
\label{eqn:sz-fz}
\end{eqnarray}
and define a magnetic field dependent spin-relaxation time scale by
the condition: $f(t_{spin}) = 0.15$. Note that the time scale
$t_{spin}$ does {\em not} have the meaning of a reciprocal relaxation
rate, since $f(t)$ is -- in general -- not an exponential function. The
choice of  $0.15$ generalizes the criterion given for $t_{low}$ at the
end of  section \ref{sec:short-and-long-time-AF}. While  the value is
arbitrary, of  course, but setting to $0.2$ or $0.3$ yields the same
qualitative picture as depicted in the inset of
Fig.~\ref{fig:finite-H-kondo}b). For very weak magnetic fields, $H\le
0.1 T_K$, the spin-relaxation time $t_{spin}(H)$ remains almost field
independent while for $H>T_K$ the relaxation time rapidly drops and
reaches $t_{low}*D = 4.0$ for (a) and  $t_{spin}*D = 9.4 $  at
$H^1_z=100T_K$. 

\begin{figure}[tb]
  \centering
  \includegraphics[width=86mm]{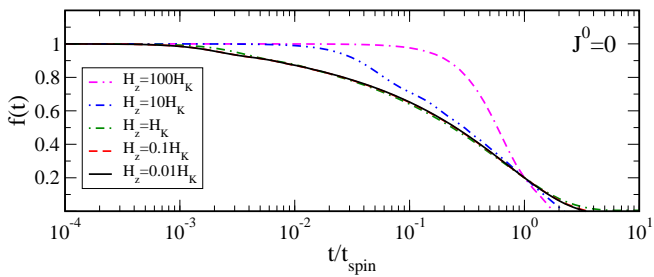}

  \caption{Data of Fig.\ \ref{fig:finite-H-kondo}, panel (a)  plotted as 
    rescaled spin evolution $f(t)$ defined in Eq.\
    \ref{eqn:sz-fz} versus $t_{spin}(H)$ for five different
    field strength and $J^0=0$. 
  }
  \label{fig:finite-H-kondo-rescale}
\end{figure}

In order to demonstrate the usefulness of the scale $t_{spin}(H)$, we
rescale the data $S_z(t)$ of Fig.\ \ref{fig:finite-H-kondo}a) in
dimensionless units using Eq.~(\ref{eqn:sz-fz}) and plot the
resulting $f(t)$ versus the dimensionless time scale $t/t_{spin}$. The
result is depicted in Fig.\ \ref{fig:finite-H-kondo-rescale}. For
magnetic fields much larger than the Kondo temperature $T_K$, the spin
relaxation is ``fast'' since the asymptotic value $S_z(\infty,H)$ is
reached rather rapidly. No Kondo correlations develop and no
universality is observed. For $H_z^1\le T_K$, however, all curves
collapse onto a universal curve $f(t/t_{spin})$ as clearly
visible in Fig.\ \ref{fig:finite-H-kondo-rescale}.

\subsection{The ferromagnetic regime}
\label{sec:kondo-model-td-nrg-ferro}

\subsubsection{Short and Long time Behavior}

\begin{figure}[tb]
  \centering
  \includegraphics[width=88mm]{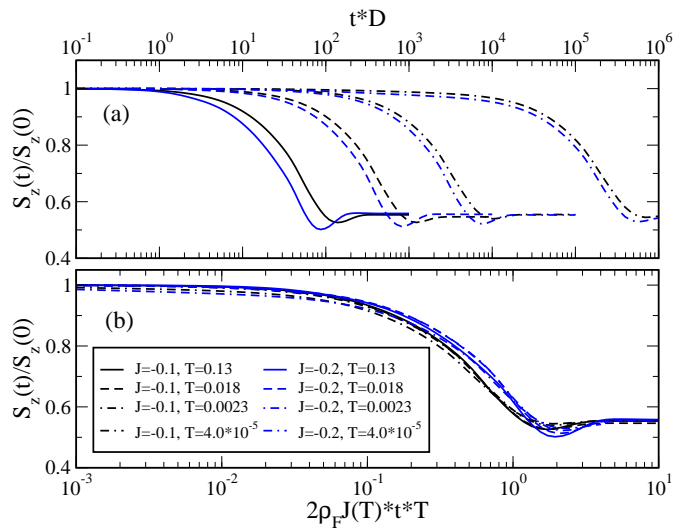}
  \caption{Normalized spin-expectation value for the isotropic Kondo
    model in the FM regime for two different values $J/D=-0.2,,-0.1$,
    and four  different temperatures 
    $T/D=0.13,0.018,0.0023, 4*10^{-5}$.  While $J=J_z=J_\perp$ is 
    kept constant, a magnetic field of $H_z/D=0.01$ is switch off at
    $t=0$. Fig (a) shows the time evolution of the normalized spin
    expectation value $S_z(t)=\expect{\hat S_z}(t)$ while in (b) the same
    data is plotted versus  the dimensionless scaling variable $x=2\rho_F
    J(T) T * t$. The result is 
    correct only for $x< 1$ since for large $x$ energy scales
    smaller than the smallest energy scale accessible for the NRG at a
    give temperature $T$ would be needed \cite{AndersSchiller2005}.
    Parameters: $\Lambda=1.5$; $N_s=1000$, $\alpha_d=0.1$, $N_z=16$.
  }
  \label{fig:FM-kondo-scen-I-scaling}
\end{figure}

As discussed in section \ref{sec:analytic}, the ferromagnetic regime
of the Kondo model is governed by the vanishing of $J_\perp$ for $T\to 0$.
For the isotropic Kondo model, the Hamiltonian flows towards the
local moment fixed point. \cite{Anderson70,Wilson75}

We start with a Kondo spin, always coupled to the bath, i.e.\
$J_\alpha=constant$ and only subject to a switched magnetic field. In
this case, the strong correlations already form at times $t<0$. 
In Fig.\ \ref{fig:FM-kondo-scen-I-scaling}a), $S_z(t)$ normalized to
$S_z(0)$ is plotted for four different temperatures and two values of 
ferromagnetic couplings $J=-0.1,-0.2$. The spin-relaxation time
increases with decreasing temperature, since the coupling $J(T)$ 
between conduction band and local spin is significantly reduced 
as predicted by the pour man scaling result (\ref{eqn:running-J}).
In Fig.\ \ref{fig:FM-kondo-scen-I-scaling}b, the same data as in
panel (a) is plotted versus the dimensionless variable
$x=2\rho_F J(T) T t$. All curves, which cover about six orders of
magnitude of time dependency, collapse very well onto a single scaling
curve. Only the effective value of $J(T)$ entered the
scaling variable $x$ in addition to explicit temperature. We did not need
to include an time dependency in  $J(T)$ as suggested
in Ref.\ \onlinecite{NordlanderEtAll1999}.

The TD-NRG does fail to predicted the vanishing of spin-expectation
value. We observe a saturation at approximately 
$S_z(0)*0.55$ which occurs at $x\approx 1$. Note, however, that  $x=1$
corresponds to $t*T= [2\rho_F J(T)]^{-1} \gg 1$ very far outside of
the validity range of the method. The NRG does not have access to
energies much lower than the temperature $T$. Nevertheless, the scaling
predictions of the TD-NRG gives very strong indications  that the
spin-relaxation rate is proportional $T$ for the isotropic ferromagnetic 
Kondo model in addition to the expected dependency of
the reduced  Kondo coupling $J(T)$. At low temperature, the effective
coupling renormalized to zero and, therefore, the perturbative
approach of  Langreth and  Wilkins\cite{LangrethWilkins1972} gives the
correct answer for the relaxation rate $\Gamma \propto (\rho_F
J)^2 k_B T$ in the long time limit. The proportionality of $\Gamma$
and the temperature stems from the fact that the number of final states for
the scattered electrons is proportional  to the thermal spread $k_B T$.

\subsubsection{Spin precession}

\begin{figure}[tbp]
  \centering

  \includegraphics[width=90mm]{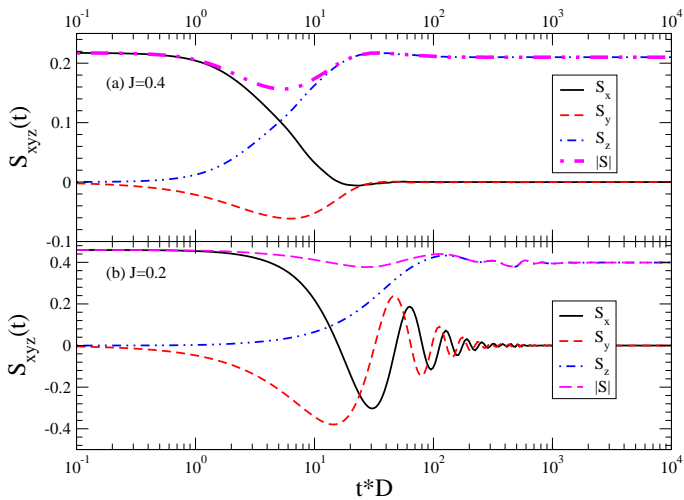}
   \caption{Spin precession of a Kondo spin coupled to the conduction
     electrons at all time $J^0_z=J^0_\perp=const$ for two different
     values of $J$. At $t=0$ the
     magnetic field $H_x/D=0.1$ is
     rotated around the $y$-axis and pointing along the
     $z$-axis, $\vec{H} = 0.1D \vec{e}_z$ for $t>0$. 
     NRG Parameters: $N_s=1000, N_z=16,\Lambda=1.6, N_{iter}=60,\alpha_d=0.1$.
   }

  \label{fig:spin-precession-AF}
\end{figure}

So far, we have only focused on the spin relaxation from a spin-polarized
state along the $z$-axis. Precession of the spin around
an  external field requires a  magnetization perpendicular to this
external field. To accomplish this,  we applied a finite magnetic
field $\vec{H}= 0.1D \vec{e}_x$ for $t<0$ to induce a magnetization in
$x$-direction. At $t=0$, the external field is rotated in
$z$-direction. In the absence of any relaxation mechanism, i.e.\ for
$J=0$, the spin would only precess in the $xy$-plane once $\vec{H}$ is
aligned along the $z$-axis. The finite Kondo interaction, however,
induces spin flips and causes spin exchange  between the  conduction
band  and the localized spin. In the thermodynamic limit and for
$t\to\infty$, the expectation value of the spin vector
$\expect{\vec{\hat S}}(t)$ will evolve into a spin aligned along the $z$-axis.

The time evolution of all spin components is displayed in Fig.\
\ref{fig:spin-precession-AF} for a magnetic field of constant strength
rotated from the $x$- to the $z$-axis at $t=0$. Since the Zeeman 
term of the Hamiltonian (\ref{eqn:zeemann}) does not commute with the
total $\hat S_z$ and only with $\vec{S}_{tot}^2$,
the TD-NRG calculations are much more time consuming.  
The black and red curve (online) show the $S_x(t)$ and $S_y(t)$ components,
respectively. As expected, the spin precesses in the $xy$-plane which
is clearly visible for $J/D=0.2$. In addition, these two components decay
with time and vanish for long times, while the $S_z(t)$ component grows
and reaches a value close to the thermodynamic expectation value with
respect to  $\H^f$. This indicates clearly that the TD-NRG can covers
more complex relaxation phenomena. Indeed the local expectation
values  tent to approach their long time thermodynamic limit,
which is a highly non-trivial feature of the method. 

\begin{figure}[tbp]
  \centering

  \includegraphics[width=88mm]{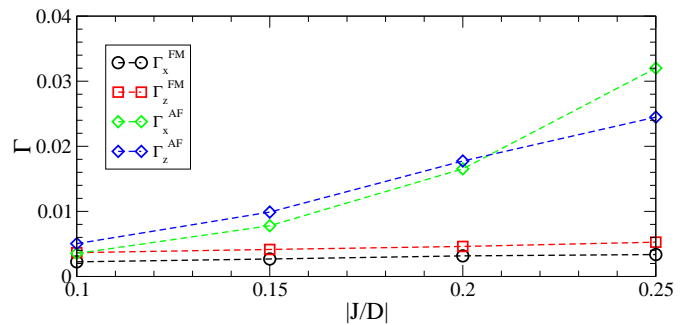}

  \caption{Relaxation rate for the $S_x$ and the $S_z$ component of
    the spin by fitting $S_x(t)=0.5\cos(\w_L t)\exp(-\Gamma_x t)$ and
    $S_z(t)=S_z(\infty)[1-\cos(\w^z_L t ) \exp(-\Gamma_z t) ]$ for
    isotropic ferromagnetic and anti-ferromagnetic coupling.
    NRG Parameters: $N_s=1000, N_z=16,\Lambda=1.8, N_{iter}=60,\alpha_d=0.1$.
  }
  \label{fig:spin-relxation-rates}
\end{figure}

For a quantitative analysis, we fitted the two spin components $S_x(t)$
and $S_z(t)$ by the following phenomenological {\em ansatz}:
$S_x(t)=S_x(0)\cos(\w_L)\exp(-\Gamma_x t)$ and
$S_z(t)=S_z(\infty)[1- \cos(\w_zt)\exp(-\Gamma_z t) ]$. 
$S_z(t)$  shows an oscillatory behavior with a frequency
$\w^z_L\approx 0.006$ for the ferromagnetic Kondo coupling and
an $\w^z_L\approx 0.02$ for the anti-ferromagnetic Kondo
coupling.  The decay of the local $S_x$ component leads to a very
small but finite magnetization in the conduction electron chain in $x$
direction. Its local component induces a spin precession of the $S_z$
which is damped by the spin-flip processes. For a chain length of
$N=60$, we estimate the average magnetization per chain link as
$S_x(0)/N=1/120\approx 0.0083$ which is a good estimate for the lower
limit  of an effective local magnetic field in $x$ direction generated by
the conduction electrons in a finite size system. In the
anti-ferromagnetic case, the coupling of the local spin to the
conduction electrons is strongly enhanced which might cause an
inhomogeneous distribution of magnetization over the Wilson chain. This
effect must vanish in the thermodynamic limit $N\to \infty,\Lambda\to 1^+$.
We find $(\Gamma_x,\Gamma_z)=(0.014,0.019)$ for $J/D=0.2$ and
$(0.12,0.113)$ for $J/D=0.4$.

In Fig.\ \ref{fig:spin-relxation-rates}, the effective spin-relaxation
rates $\Gamma_x$ and $\Gamma_z$ are displayed as function of $|J/D|$
for an initially decoupled local spin, i.e.\ $J(t<0)=0$.  With
the exception of $J/D=0.25$, $\Gamma_x$ is smaller than $\Gamma_z$ for
anti-ferromagnetic coupling  while the transversal relaxation is
always slower than the longitudinal for ferromagnetic  coupling.
However, the absolute values strongly depend on the fitting 
functions for $S_x(t)$ and $S_z(t)$, so that the relaxation rates are
roughly equal for all directions. Note that the usual dephasing time
$T_2$ in NMR and ESR experiments originates from an average over
signals stemming from many spins subject to local fluctuations of the
externally applied field. This inhomogeneities yields the ESR
condition $1/T_2> 1/T_1$ which does not necessarily hold for the
single spin decay investigated here.

\section{Summary and discussion}

\subsection{Discussion}

We have presented a powerful novel approach to the nonequilibrium dynamics
of quantum impurity systems  based upon Wilson's NRG. Its virtue is
based on three key points: (i) all states
of the Wilson chain are used for the time-evaluation of local
operator, (ii) time expectation values can be expressed by a summation
of all NRG iterations, where at each iteration the matrix elements of
any local operator is weight by the reduced density matrix which
contain information about dissipation and decoherence and (iii) the
continuum limit of the bath is mimicked by using the $z$-trick where the
true time evolution is obtained by averaging over a different bath
discretization. By its nature, it is applicable to arbitrary
temperatures,  and allows for the study of the temperature evolution
of real-time dynamics.

We have previously established the accuracy of the
approach on all time scales up to $O(t \, T) $ by comparison with exact
solution of the RLM \cite{AndersSchiller2005}. In this paper, we
additionally benchmarked our method to the exact analytical solution
of the decay of the off-diagonal matrix element $\rho_{01}$ of the
local density matrix which is taken as measure for decoherence of a
spin state. This is the first application of the TD-NRG to bosonic
baths. In this case, the TD-NRG is much less susceptible to
discretization errors than for fermionic baths. This
behavior roots in the extended number of bath states, in principle
infinite, which are available for distributing energy and phase at
each energy scale $D_N$. From this observation, we conjecture that the
TD-NRG will be even more suitable from bosonic baths than for
fermionic baths. A detailed study of the real-time dynamics of the
spin-boson model in the subohmic regime will be published
soon. Combining fermionic and bosonic baths\cite{GlossopIngersent2005}
will open new possibilities  for the description of spin decay of
ultra-small quantum dots  coupled to the leads as well as charge noise.

We have investigated the spin decay of the Kondo model in the ferromagnetic
and in the antiferromagnetic regime. For the ferromagnetic regime, we
identified a universal dimensionless time scale $x=2\rho_F J(T)T t$ for
the transient behavior of the spin decay.  In the antiferromagnetic
regime, the  spin response at short time scales follows excellently
the presented 
analytical perturbative solution which is valid up the second order in 
the coupling constant. At long times, the Kondo correlations dominate,
and the rescaled curves appear to collapse onto a universal curve for
$t/t_K\to \infty$ as predicted by conformal field theory
\cite{LesageSaleur1998}. At short and intermediate times $t\approx
t_K$, the anisotropy of $J_z$ and $J_\perp$ influences the time
evolution and no universality is found as a one parameter scaling for
different values of $J_z/J_\perp$. This reflects the different role of
$J_z$ and $J_\perp$ in the spin-relaxation process. The spin
relaxation is governed by the spin-flip term $J_\perp$ and not by
$J_z$ which only enters through renormalization in higher order
processes into $J_\perp$. Only at $T=0$ and at infinitely long times,
the anisotropy Kondo model becomes isotropic in the strong coupling
fixed point. Here, we expect the validity of universal scaling.

\subsection{Outlook}

The time-dependent NRG approach will open new doors to our
understanding of a new class of nonequilibrium problems whose
real-time dynamics is governed by a strong entanglement of the environment 
with the quantum-impurity states. The exact analytical solution of
decoherence of a pure quantum states explicitly shows that the true
dynamics is not simply given by an exponential decay rate $\Gamma$ but
a function $\Gamma(t)$. The reproduction of these non-trivial
analytical results of the spin-boson model with all details reveals 
already the strong potential of our method. The application 
to charge-transfer reactions in biological
systems\cite{TornowBulla2005} influenced by the molecular vibrations
is subject of an ongoing research project. We hope that this 
will ultimately lead to a better understanding how secondary and
tertiary molecular structures of protein molecules influence the
chemistry and, therefore, the functionality of such 
proteins. It was pointed out by Schulten
\cite{Schulten2000} that the chemical structure of reaction centers in
protein molecules by itself does not give us the desired
understanding why thermal fluctuations do no harm the deterministic
outcome of the complex cellular chemistry which is the basis of 
existence of living organisms.

We have benefited from discussions with G.~Alber,
R.~Bulla, G.~Czycholl, T.~Costi, S.~Kehrein, A.~Hewson,
E.~Lebanon, A.\ Rosch, K.~Sch\"onhammer, S.~Tornow, D.~Vollhardt,
M.~Vojta and A. Weichselbaum.
F.B.A. acknowledges funding of the NIC,
Forschungszentrum J\"ulich under project no. HHB000,
and DFG funding under project AN 275/5-1. A.S. was
supported in part by the Centers of Excellence Program
of the Israel Science Foundation.


\appendix

\section{Calculation of the overlap matrix elements}
\label{app:overlap-matrix}

In this appendix, we  describe in detail the calculation of the overlap
matrix elements of Eq.~(\ref{eq:S_kr-def}). We need the matrix element
\begin{eqnarray}
 S_{r,s}(m) &=&   \bra{s;m} r;m \rangle
\end{eqnarray}
in order to calculate the reduced density matrix in basis of the
Hamiltonian $\H^f_m$ needed in  Eq.\ (\ref{eqn:rho-pp-m}). Here,
$\ket{s;m}$ is an eigenstate of  $\H_m^i$ and $\ket{r;m}$ an eigenstate
of $\H_m^f$ and the environment variable $e$ has been droped for
simplicity.  Independent of the dynamics of $\H_m^{i/f}$, the original basis
of the Hamiltonian matrices $\mat{H}_0^i$ and $\mat{H}_0^f$ is
identical prior to diagonalization. Therefore, the overlap is given by
\begin{eqnarray}
\label{eqn:a2}
  S_{\alpha_{imp},\alpha_{imp}'} &=&  \bra{\alpha_{imp};0}
  \alpha_{imp}',0 \rangle  = \delta_{\alpha_{imp},\alpha_{imp}'}
\end{eqnarray}
where $\alpha_{imp}$ labels the states on the impurity. Let us assume,
we know all matrix elements  $S_{r,s}(m)$ at iteration $m$. Then, the
NRG recursion relation adds the {\em same} degrees of freedom
$\alpha_{m+1}$ to the new chain of length $m$, independent of the
dynamics of $\H_{m+1}^{i/f}$.  Since the new eigenstates can be
expanded into
\begin{eqnarray}
  \ket{r;m+1} &=& \sum_{\alpha_{m+1},k} P_{r,k}^{f}[\alpha_{m+1}]
  \ket{k,\alpha_{m+1}} \\
  \ket{s;m+1} &=& \sum_{\alpha_{m+1},k} P_{s,k'}^{i}[\alpha_{m+1}]
  \ket{k',\alpha_{m+1}}
\komma
\end{eqnarray}
the matrix element  $S_{r',s'}(m+1)$ is given by
\begin{eqnarray}
  S_{r',s'}(m+1) &=& \bra{s';m+1} r'; m+1\rangle
\\
&=&
\sum_{k,k'}\sum_{\alpha_{m+1},\alpha_{m+1}'}
 P_{r',k}^{f}[\alpha_{m+1}]
 P_{s',k'}^{i}[\alpha_{m+1}']
\nonumber
\\
&&
\times
\bra{k,\alpha_{m+1}} k',\alpha_{m+1}' \rangle
\nonumber
\\
&=&
\sum_{k,k'}\sum_{\alpha_{m+1}}
 P_{r',k}^{f}[\alpha_{m+1}]
\nonumber
\\
&& 
\times 
 P_{s',k'}^{i}[\alpha_{m+1}] S_{k,k'}(m)
\punkt
\nonumber
\label{eqn:a5}
\end{eqnarray}
In combination with the initial condition (\ref{eqn:a2}), Eq.\
(\ref{eqn:a5}) defines a recursion relation from which all overlap
matrix elements are  obtained.


\end{document}